\documentclass[aps,twocolumn,prb]{revtex4-1}
\usepackage{amsfonts}
\usepackage{amsmath}
\usepackage{amssymb}
\usepackage{graphicx}

\begin{document}

\title{Bond distortion effects and electric orders
in spiral multiferroic magnets}
\author{Hong-Bo Chen, Yi Zhou and You-Quan Li}
\date{\today}
\affiliation{Zhejiang Institute of Modern Physics and Department of Physics, Zhejiang
University, Hangzhou 310027, People's Republic of China}

\begin{abstract}
We study in this paper bond distortion effect on electric polarization in
spiral multiferroic magnets based on cluster and chain models. The bond
distortion break inversion symmetry and modify the $d$-$p$ hybridization.
Consequently, it will affect electric polarization which can be divided into
spin-current part and lattice-mediated part. The spin-current polarization
can be written in terms of $\vec{e}_{i,j}\times(\vec{e}_{i}\times\vec{e}%
_{j}) $ and the lattice-mediated polarization exists only when the M-O-M
bond is distorted. The electric polarization for three-atom M-O-M and
four-atom M-O$_{2}$-M clusters is calculated. We also study possible
electric ordering in three kinds of chains made of different clusters.
We apply our theory to multiferroics cuprates and find that the results
are in agreement with experimental observations.
\end{abstract}

\pacs{}
\maketitle

\section{Introduction}

Magnetoelectric multiferroics, where magnetic and electric ordering
coexisting in a single compound, has attracted much interest in the last
decades due to their potential for novel physics and technological
applications \cite{Fiebig}. However, the initial observations have shown
that materials with coexisting ferroelectric and magnetic orders are rare in
nature and the observed magnetoelectric coupling was rather
weak\cite{Hill}. Therefore, the search for new gigantic magnetoelectric multiferroic
materials is a great challenge. As an important progress in multiferroics,
Kimura \textit{et al}. \cite{Kimura} found giant magnetoelectric effects in
the perovskite manganite TbMnO$_{3}$ in 2003. The ferroelectric transition
at $T_{c}=28$ K in TbMnO$_{3}$ was ascribed to the emergence of a nontrivial
magnetic order, i.e., with a spiral spin structure \cite{Kenzelmann}. This
finding revived the interests in multiferroic behavior and lead to the
discovery of a relatively large number of spin-driven multiferroic materials
\cite{Chenong,Khomskii,Liu09}.

One of the fundamental issues in multiferroic physics is to clarify the
origin of the magnetoelectric coupling in such new materials. Currently,
although several models have been proposed both microscopically \cite%
{Katsura05,Jia2006,Jia2007,JPHu,Hu07,Hu08,Hu10,Dagotto06,Mochizuki-prl10,Mochizuki-prb11} and
phenomenologically \cite{Mostovoy}, the microscopic mechanism of the
magnetoelectric coupling is still a controversial and unresolved topic. One
of the most prevailing mechanisms is the spin-current model proposed by
Katsura \textit{et al}. \cite{Katsura05} for helical magnets. They studied a
three-atom M-O-M (M denotes a transition metal ion and O is for an oxygen
ion) cluster where O locates in the inversion center of the cluster, say,
the mid point between two M ions. They pointed out that spontaneous electric
polarization in the spiral magnetic order arises from a distortion of
electronic density even without ionic or atomic displacements. And the
polarization can be expressed in the form
\begin{equation}
\vec{P}\propto\vec{e}_{i,j}\times(\vec{e}_{i}\times\vec{e}_{j}),  \label{KNB}
\end{equation}
where $\vec{e}_{i,j}$ denotes the bond direction connecting the two neighbor
spin moments $\vec{S}_{i}$ and $\vec{S}_{j}$ along directions $\vec{e}_{i}$
and $\vec{e}_{j}$ on the sites $i$ and $j$ respectively. The electric
polarization is closely associated with the nonzero spin current $\vec{S}%
_{i}\times\vec{S}_{j}$ between two neighboring noncollinear spins.
Furthermore, Jia \textit{et al}. \cite{Jia2006,Jia2007} presented a more
sophisticated calculation on the spin-current model with realistic
considerations. Their results showed that the spin-current model is able to
explain many experimental data at least semiquantitatively. On the other
hand, Sergienko and Dagotto \cite{Dagotto06} argued that the ferroelectric
polarization is closely related to the inverse Dzyaloshinskii-Moriya(DM)
coupling which breaks inversion symmetry and obtained similar expression for
electric polarization as in (\ref{KNB}). Interestingly, the spin-orbit
interaction plays a crucial role in both theories. The role of spin-orbit
interaction was also confirmed by Mostovoy based on the phenomenological
theory\cite{Mostovoy}. Experimentally, the spin-current mechanism was
demonstrated directly by controlling the spin-helicity vector in TbMnO$_{3}$
with an external electric field \cite{Yamasaki}. The spin-current mechanism
has also been applied to several other spiral multiferroic
materials, such as DyMnO$_{3}$ \cite{DMO04}, MnWO$_{4}$ \cite{MWO06}, Ni$%
_{3} $V$_{2}$O$_{8}$ \cite{NVO05}, CoCr$_{2}$O$_{4}$ \cite{CoCrO06}, LiCu$%
_{2}$O$_{2}$ \cite{LCuO07,LCuO08} and LiCuVO$_{4}$ \cite{LCuVO07,LCuVO08-Yasui,LCuVO08-schrettle,LCuVO11}. Beside
these successful applications, it should be noted that the magnitude of
electric polarization is very sensitive to the location of Fermi energy in
existing spin-current models. When the Fermi energy locates between two
hybridized levels, which refers to ``single hole" in Ref.\cite{Katsura05}, $%
P\propto V/\Delta$, where $V$ is the hybridization between transition metal $%
d $-orbitals and oxygen $p$-orbitals and $\Delta$ is the energy difference
between $d$-orbitals and $p$-orbitals. Whereas, when the Fermi energy locates
outside two hybridized levels, which refers to ``double hole" in Ref.\cite%
{Katsura05}, $P\propto (V/\Delta)^3$, which is significantly different from
``single hole" situation.

In real multiferroics, their fascinating properties are attributed to the
competition among charge, spin, orbital, and lattice degrees of freedom.
Actually, the bond-distortion is inevitably present in realistic transition
metal oxides due to structure distortion \cite{Kimura03,Arima}. Thus, the
inversion symmetry breaks and the oxygen atom is away from the mid point
between two M atoms. Therefore, the bond-distortion effects should be taken
into account in the calculation for the electric polarization. In existing
spin-current model \cite{Katsura05,Jia2006,Jia2007}, the bond-distortion
effect is overlooked. As a result that we shall show in this paper, the
calculated electric polarization is underestimated. A recent theoretical
study on a toy model for the ferroelectricity of a two-dimensional cluster
\cite{Hu08} suggests that the bond-bending may be important for the enhancement
of the ferroelectricity due to orbital hybridization. On the other hand,
Moskvin \textit{et al}. \cite{Moskvin08} argued that the cluster model in
Ref. \cite{Katsura05} is oversimplified to account for LiCu$_{2}$O$_{2}$ and
LiCuVO$_{4}$ in which the realistic geometry configuration of the oxygen
atoms was not considered. All of these facts motivate the present work to
explore the effect of bond-distortion on the electric polarization in the
spiral multiferroic magnets in the framework of spin-current model.

The organization of this paper is as follows. In Section \ref{Sec:model}, we
perform quantum chemistry analysis on relevant $d$-orbitals and construct
cluster models with two M atoms and bridge oxygen atoms. Then we calculate
lowest lying eigenstates for these cluster models. In Section \ref%
{Sec:P-Cluster}, we calculate the electric polarization for different
clusters. In Section \ref{Sec:chain-models}, we discuss three kinds of chain
models formed by different clusters and possible ordering of electric
polarization. In section \ref{Sec:discussion}, we discuss the relation between
lattice-driven polarization and magnetic-driven polarization
and apply our theory to multiferroic copper oxides.
Section \ref{Sec:conclusion} is devoted to summary.

\section{Cluster models}

\label{Sec:model}

\begin{figure}[ptbh]
\centering
\includegraphics[width=6.0cm]{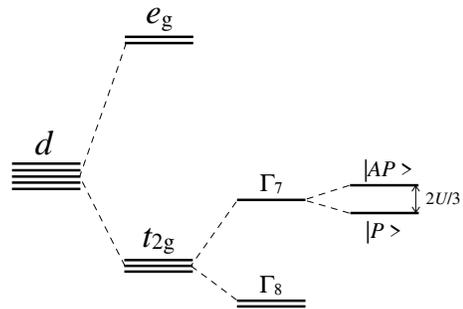}
\caption{ Relevant electronic states at transition metal atoms. Five fold
degenerate $d$-orbitals will split into two fold $e_{g}$ and three fold $%
t_{2g}$ levels. The spin-orbit interaction $\protect\lambda$ splits $t_{2g}$
levels to a doublet $\Gamma_7$ and a quartet $\Gamma_8$. In the presence of
effective exchange $U$ between $\Gamma_7$ electrons and local magnetic
moment, $\Gamma_7$ will be further split into spin parallel state $%
\left\vert P\right\rangle$ and antiparallel state $\left\vert
AP\right\rangle $. }
\label{fig:multiplets}
\end{figure}

We begin with quantum chemistry consideration. As is well known, when the $%
3d $ transition metal atoms are placed in an octahedral crystal field, the
five-fold degenerate $d$ levels will split into $e_{g}$ and $t_{2g}$ levels.
Incorporating the on-site spin-orbit interaction $\lambda$, the $e_{g}$
orbitals will not be influenced but the $t_{2g}$ manifold will further split
into a doublet $\Gamma_{7}$ and a quartet $\Gamma_{8}$ (see Fig.\ref{fig:multiplets}). For simplicity and
following Katsura \textit{et al}\cite{Katsura05}, we will adopt the
following $\Gamma_{7}$ states as the ground state manifold and drop the
quartet,
\begin{eqnarray}
\left\vert a\right\rangle & =\frac{1}{\sqrt{3}}\left( \left\vert
d_{xy,\uparrow}\right\rangle +\left\vert d_{yz,\downarrow}\right\rangle
+i\left\vert d_{zx,\downarrow}\right\rangle \right) ,  \notag \\
\left\vert b\right\rangle & =\frac{1}{\sqrt{3}}\left( \left\vert
d_{xy,\downarrow}\right\rangle -\left\vert d_{yz,\uparrow}\right\rangle
+i\left\vert d_{zx,\uparrow}\right\rangle \right) ,  \label{Eq.ab}
\end{eqnarray}
by assuming the spin-orbit coupling $\lambda$ is the largest energy scale of
the problem.

As in Ref.\cite{Katsura05}, we further invoke the effective exchange
interaction between $\Gamma_7$ electrons and local magnetic moment governed
by the following Hamiltonian,
\begin{equation}  \label{Eq.HU}
H_{U}=-U\sum_{j}\vec{e}_{j}\cdot\vec{S}_{j},
\end{equation}
where $\vec{e}_{j}$ $=(\cos\phi_{j}\sin\theta_{j}$, $\sin\phi_{j}\sin
\theta_{j}$, $\cos\theta_{j})$ is the unit vector of the local magnetic
moment from the transition metal atom $\mathrm{M}$ at $j$-th site, $\vec{S}%
_{j}$ the total spin operator of the $d$-orbital electrons, $U$ the
effective exchange interaction which is of the order of Coulomb interaction
and Hund's coupling energy\cite{Katsura05,Jia2006,Jia2007}. Then $H_U$ will
further split the $\Gamma_7$ doublet into spin parallel state $\left\vert
P\right\rangle$ and anti-parallel state $\left\vert AP\right\rangle$,
\begin{subequations}
\label{Eq.PAP}
\begin{eqnarray}
\left\vert P_{j}\right\rangle & = & \sin\frac{\theta_{j}}{2}\left\vert
a\right\rangle +e^{i\phi_{j}}\cos\frac{\theta_{j}}{2}\left\vert
b\right\rangle ,  \label{Eq.P} \\
\left\vert AP_{j}\right\rangle & = & \cos\frac{\theta_{j}}{2}\left\vert
a\right\rangle -e^{i\phi_{j}}\sin\frac{\theta_{j}}{2}\left\vert
b\right\rangle,  \label{Eq.AP}
\end{eqnarray}
where $|P_{j}\rangle$ and $|AP_{j}\rangle$ indicate the spin state parallel
and anti-parallel to the unit vector ${\vec{e}}_{j}$, and the corresponding
eigenvalues are $E_{|P_{j}\rangle}=-\frac{U}{3}$ and $E_{|AP_{j}\rangle}=%
\frac{U}{3}$, respectively. Note that we may further write $|P_{j}\rangle$
and $|AP_{j}\rangle$ in terms of $t_{2g}$ states,
\end{subequations}
\begin{subequations} 
\label{Eq.PAPt2g}
\begin{equation}  \label{Eq.PA}
|P_{j}\rangle=\sum_{\mu\sigma}A_{(j)}^{\mu\sigma}|d_{\mu\sigma}^{(j)}\rangle,
\end{equation}
and
\begin{equation}  \label{Eq.APB}
|AP_{j}\rangle=\sum_{\mu\sigma}B_{(j)}^{\mu\sigma}|d_{\mu\sigma}^{(j)}%
\rangle,
\end{equation}
where $\mu=xy,yz,zx$, $\sigma=\uparrow,\downarrow$, $A_{(j)}^{\mu\sigma}$
and $B_{(j)}^{\mu\sigma}$ are coefficients obtained from combining Eqs.(\ref%
{Eq.PAP}) and (\ref{Eq.PAPt2g}) which depend on angles $\theta_{j}$ and $%
\phi_{j}$ (See Appendix \ref{App1} for details). Hereafter, we assume that $%
\lambda$ and $U$ are much larger than other relevant energy scales so that
there is only one relevant state in each M atoms, say, $\left\vert
P\right\rangle$. We also assume that there is a spiral magnetic order
relating to the transition metal atoms.

Now we are in the position to construct cluster models by including bridge
oxygen atoms between two M atoms. Taking the hole picture, we assume that
the oxygen $p$-orbital has energy level $E_p$ which is above $E_{|P\rangle}$
with a difference $\Delta$ as shown in Fig.\ref{fig:eigenstates}, i.e. $%
E_p=E_{|P\rangle}+\Delta$. The hybridization between spin parallel state $%
|P\rangle$ and oxygen $p$-orbitals results in bonding and anti-bonding
states with lower and higher energy respectively. In such a M-O-M cluster
(see Fig.\ref{fig:eigenstates}), left and right bonding states will further
hybridize with each other and lead to two lowest energy levels $E_1$ and $%
E_2 $. When the Fermi energy $E_F$ locates between $E_1$ and $E_2$, it is
called ``single hole" in Ref.\cite{Katsura05}. Whereas, it is called
``double hole" when $E_F$ is above $E_2$ but below other levels.

\begin{figure}[ptbh]
\centering
\includegraphics[width=8.5cm]{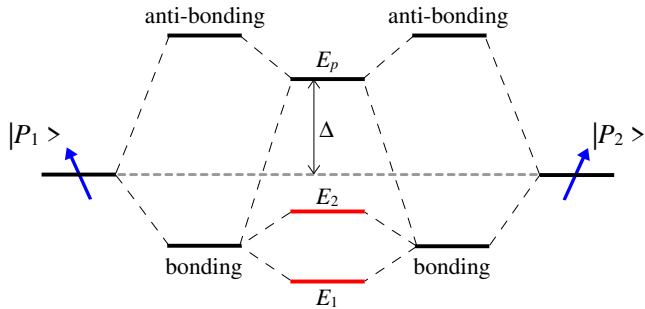}
\caption{Energy levels demonstration for a M-O-M cluster.}\label{fig:eigenstates}
\end{figure}

The hybridization between left and right bonding states will also lead to
distorted electronic density and thereby possible electric polarization in
such a M-O-M cluster. We may further consider more bridge oxygen atoms
between two M atoms and derive similar electronic density distortion and
electric polarization. We shall study three-atom M-O-M cluster and four-atom
M-O$_2$-M cluster in the remained part in this section and calculate
electric polarization for different kinds of chains formed by three-atom or
four-atom clusters in next section.

\subsection{Three-atom M-O-M cluster}

Firstly we consider the three-atom M-O-M cluster model shown in Fig.\ref%
{fig:model}. M$_{1}$ and M$_{2}$ refer to two transition metal atoms and the
intermediate oxygen atom (O) deviates from its centrosymmetric position on M$%
_{1}$-M$_{2}$ bond. To be simple, we assume that the three atoms of the
cluster are restricted within the $xy$-plane, the M$_{1}$-M$_{2}$ bond is
along the $x$-axis, and the lengths of two M-O bonds are equal to each
other. Note that Katsura \textit{et al}. \cite{Katsura05} considered a
centrosymmetric three-atom model in which the oxygen atom locates at the mid
point between two M atoms.

\begin{figure}[ptbh]
\centering
\includegraphics[width=5cm]{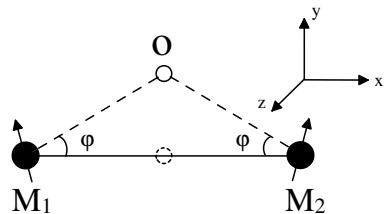}
\caption{Three atom M-O-M cluster model. M$_{1}$ and M$_{2}$ denote two
transition metal ions with noncollinear magnetic moments. O represents a
bridge oxygen atom which deviates from its centrosymmetric position (denoted
by dash circle) with a bond-bending angle $\protect\varphi$. The arrows on $%
\mathrm{M}_{1}$ and $\mathrm{M}_{2}$ indicate the spin direction. The three
atoms of the cluster are assumed lying in the $xy$ plane.}
\label{fig:model}
\end{figure}

Upon the above quantum chemistry analysis, we use the following Hamiltonian
to describe the on-site energy for oxygen $p$-orbitals and spin parallel
states,
\end{subequations}
\begin{equation}
H_{\Delta}=\sum_{j=1,2}E_{\vert P\rangle} c_{\vert
P_j\rangle}^{\dag}c_{\vert P_j\rangle} +\sum_{\mu,\sigma}(E_{\vert
P\rangle}+\Delta) p_{\mu,\sigma}^{\dag}p_{\mu,\sigma},  \label{Eq.HD}
\end{equation}
where $p_{\mu,\sigma}$, $(\mu=x,y,z)$ are annihilation operators for oxygen $%
p$-electrons, $\sigma$ is spin index, $c_{\vert P_j\rangle}$ is the
annihilation operator for spin parallel state at site $j$.

The hybridization between $p$-orbitals and spin parallel states is governed
by the hopping Hamiltonian,
\begin{equation}
H_{t}=H_{t}^{1-m}+H_{t}^{2-m}+\mathrm{H.c.,}  \label{Eq.Ht}
\end{equation}
where $H_{t}^{1-m}$ and $H_{t}^{2-m}$ indicate the hopping processes between
the oxygen atom and M$_{1}$ and M$_{2}$ respectively. We then write down $%
H_{t}^{1-m}$ and $H_{t}^{2-m}$ in terms of nonvanishing hopping integrals
explicitly,
\begin{align}
H_{t}^{1-m} & =\sum_{\sigma}\Big (V_{1}p_{x,\sigma}^{\dag}d_{xy,\sigma
}^{(1)}+V_{2}p_{y,\sigma}^{\dag}d_{xy,\sigma}^{(1)}  \notag \\
& \ \ \ \ \ \ \ \ \ \ \ \
+V_{3}p_{z,\sigma}^{\dag}d_{zx,\sigma}^{(1)}+V_{4}p_{z,\sigma}^{\dag}d_{yz,%
\sigma}^{(1)}\Big ),  \label{Eq.Ht1m} \\
H_{t}^{2-m} & =\sum_{\sigma}\Big (V_{1}p_{x,\sigma}^{\dag}d_{xy,\sigma
}^{(2)}-V_{2}p_{y,\sigma}^{\dag}d_{xy,\sigma}^{(2)}  \notag \\
& \ \ \ \ \ \ \ \ \ \ \ \ \
-V_{3}p_{z,\sigma}^{\dag}d_{zx,\sigma}^{(2)}+V_{4}p_{z,\sigma}^{\dag}d_{yz,%
\sigma}^{(2)}\Big ),  \label{Eq.Ht2m}
\end{align}
where $d_{\mu,\sigma}^{(j)}$ $(\mu=xy,yz,zx)$ are annihilation operators for
$d$ electrons of the transition metal atoms. The superscript $j=1,2$ in $%
d_{\mu,\sigma}^{(j)}$ denote the site of magnetic transition metal atoms as
shown in Fig.\ref{fig:model}. The integrals $V_{n}$ ($n=1,2,3,4$) between $d$
and $p$ orbitals are given according to Slater-Koster's rules \cite{Slater}
as follows,
\begin{align}
V_{1} & =\sin\varphi\left( \sqrt{3}\cos^{2}\varphi
t_{pd\sigma}+(1-2\cos^{2}\varphi)t_{pd\pi}\right) ,  \notag \\
V_{2} & =\cos\varphi\left( \sqrt{3}\sin^{2}\varphi
t_{pd\sigma}+(1-2\sin^{2}\varphi)t_{pd\pi}\right) ,  \notag \\
V_{3} & =\cos\varphi t_{pd\pi},  \notag \\
V_{4} & =\sin\varphi t_{pd\pi}.  \label{hopping}
\end{align}
Here $\varphi$ is the angle between M$_{1}$-M$_{2}$ bond (i.e., along $x$%
-axis) and M$_{1(2)}$-O bond, as shown in Fig. \ref{fig:model}. $%
t_{pd\sigma} $ and $t_{pd\pi}$ are hybridization integrals corresponding to $%
\sigma$ and $\pi$ bonding between $p$ and $d$ orbitals respectively. In
contrast to the hopping Hamiltonian derived by Katsura \textit{et al}. in
Ref. \cite{Katsura05} 
where only $t_{pd\pi}$ is present due to inversion symmetry, there are four
relevant and independent hopping integrals $V_{n},n=1,2,3,4$. Because more
active orbitals are involved in our model due to the bond-distortion, such
as $p_{x}$ and $d_{yz}$ orbitals. Thus, the Hilbert space in the present
model is enlarged. The basis of the our cluster system contains $%
|P_{j}\rangle$, $(j=1,2)$ and $p_{\mu,\sigma}$ $(\mu=x,y,z,\
\sigma=\uparrow,\downarrow)$ and an eigenstate is a linear combination of
these states. We can easily recover centrosymmetric cluster model in Ref.%
\cite{Katsura05} by setting $\varphi=0$, which leads to $V_1=V_4=0$ and $%
V_2=V_3$.

We proceed to calculate the eigenstates of the system by assuming $%
V\ll\Delta $ and treating $H_{t}$ as a perturbation to $H_{\Delta}$. To the
second order perturbation\cite{Landau}, the two lowest lying states are
given by energy levels
\begin{equation}  \label{E3atom}
E_{1(2)} = 2(C \mp \left\vert B\right\vert),
\end{equation}
with corresponding eigenvectors $\left\vert 1\right\rangle$ and $\left\vert
2\right\rangle$ given in Appendix \ref{App3}, where the parameters $C$ and $%
B $ are as follows,
\begin{eqnarray}  \label{CB}
C & = & -\frac{V_{1}^{2}+V_{2}^{2}+V_{3}^{2}+V_{4}^{2}}{3\Delta},  \notag \\
B & = & A_{1}\alpha+A_{2}\beta,
\end{eqnarray}
with
\begin{eqnarray}  \label{A12}
A_{1} & = & -\frac{V_{1}^{2}-V_{2}^{2}-V_{3}^{2}+V_{4}^{2}}{3\Delta},  \notag
\\
A_{2} & = & \frac{2V_{3}V_{4}}{3\Delta},
\end{eqnarray}
and
\begin{eqnarray}  \label{alphabeta}
\alpha & = &\sin\frac{\theta_{1}}{2}\sin\frac{\theta_{2}}{2}+\cos\frac{%
\theta_{1}}{2}\cos\frac{\theta_{2}} {2}e^{-i(\phi_{1}-\phi_{2})},  \notag \\
\beta & = & i[\sin\frac{\theta_{1}}{2}\sin \frac{\theta_{2}}{2}-\cos\frac{%
\theta_{1}}{2}\cos\frac{\theta_{2}} {2}e^{-i(\phi_{1}-\phi_{2})}].
\end{eqnarray}
The angles $(\theta_{1},\phi_{1})$ and $(\theta_{2},\phi_{2})$ specify the
direction of local magnetic moments in transition metal atom M$_1$ and M$_2$
respectively. For a centrosymmetric cluster with $\varphi=0$, we have $A_2=0$
and $C=-A_1$.

\subsection{Four-atom M-O$_2$-M cluster}

\label{subsec:four-site}

Secondly, we consider the four-atom M-O$_2$-M cluster illustrated in Fig.\ref%
{fig:symmetric}. In this situation, there are two oxygen atoms
symmetrically deviated away from the M$_{1}$-M$_{2}$ bond with bending angle
$\varphi$. Here, we also assume that the atoms of the cluster are restricted
within the $xy$ plane and the M$_{1}$-M$_{2}$ bond is along the $x$-axis for
convenience.

\begin{figure}[hpbt]
\centering
\includegraphics[width=6cm]{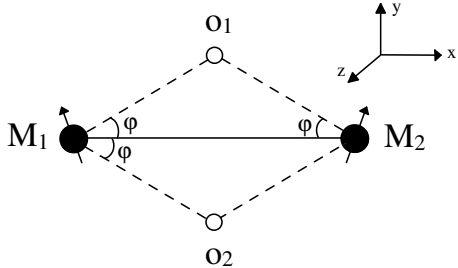}
\caption{Four-atom M$_{1}$-O$_{1(2)}$-M$_{2}$ cluster. Two bridge oxygen
atoms O$_{1}$ and O$_{2}$ symmetrically deviates from M$_{1}$-M$_{2}$ bond
with a bond-distortion angle $\protect\varphi$. The arrows on $\mathrm{M}%
_{1} $ and $\mathrm{M}_{2}$ indicate the directions of spins. The four atoms
of this cluster lie within the $xy$ plane and M$_{1}$-M$_{2}$ bond lies
along the $x$-axis.}
\label{fig:symmetric}
\end{figure}

The on-site energy in Eq.(\ref{Eq.HD}) is then modified as
\begin{equation}
H_{\Delta}=\sum_{j=1,2}E_{\vert P\rangle} c_{\vert
P_j\rangle}^{\dag}c_{\vert P_j\rangle} + \sum_{i,\mu,\sigma}(E_{\vert
P\rangle}+\Delta) p_{\mu,\sigma}^{\dag(i)}p^{(i)}_{\mu,\sigma},
\label{Eq.HD4}
\end{equation}
where the superscript $i=1,2$ in $p^{(i)}_{\mu,\sigma}$, $(\mu=x,y,z)$
denote the site number for oxygen atoms O$_1$ and O$_2$ shown in Fig.\ref%
{fig:symmetric}. Note that we assume that $p$ orbitals in the two oxygen
atoms O$_1$ and O$_2$ has the same energy levels, say, $E_{p_1}=E_{p_2}=E_{%
\vert P\rangle}+\Delta$, due to the symmetric deviation of two oxygen atoms.

Accordingly, we shall replace hopping terms $H_{t}^{1-m}$ and $H_{t}^{2-m}$
in Eq. (\ref{Eq.Ht1m}) and Eq. (\ref{Eq.Ht2m}) by
\begin{align}
H_{t}^{1-m} & =\sum_{\sigma}\Big (V_{1}p_{x,\sigma}^{\dag(1)}d_{xy,\sigma
}^{(1)}+V_{2}p_{y,\sigma}^{\dag(1)}d_{xy,\sigma}^{(1)}+V_{3}p_{z,\sigma}^{%
\dag(1)}d_{zx,\sigma}^{(1)}  \notag \\
& \ \ \ \ \ \
+V_{4}p_{z,\sigma}^{\dag(1)}d_{yz,\sigma}^{(1)}-V_{1}p_{x,\sigma}^{%
\dag(2)}d_{xy,\sigma}^{(1)}+V_{2}p_{y,\sigma}^{\dag (2)}d_{xy,\sigma}^{(1)}
\notag \\
& \ \ \ \ \ \
+V_{3}p_{z,\sigma}^{\dag(2)}d_{zx,\sigma}^{(1)}-V_{4}p_{z,\sigma}^{%
\dag(2)}d_{yz,\sigma}^{(1)}\Big ),  \label{H1t4}
\end{align}
and
\begin{align}
H_{t}^{2-m} & =\sum_{\sigma}\Big (V_{1}p_{x,\sigma}^{\dag(1)}d_{xy,\sigma
}^{(2)}-V_{2}p_{y,\sigma}^{\dag(1)}d_{xy,\sigma}^{(2)}-V_{3}p_{z,\sigma}^{%
\dag(1)}d_{zx,\sigma}^{(2)}  \notag \\
& \ \ \ \ \ \
+V_{4}p_{z,\sigma}^{\dag(1)}d_{yz,\sigma}^{(2)}-V_{1}p_{x,\sigma}^{%
\dag(2)}d_{xy,\sigma}^{(2)}-V_{2}p_{y,\sigma}^{\dag (2)}d_{xy,\sigma}^{(2)}
\notag \\
& \ \ \ \ \ \
-V_{3}p_{z,\sigma}^{\dag(2)}d_{zx,\sigma}^{(2)}-V_{4}p_{z,\sigma}^{%
\dag(2)}d_{yz,\sigma}^{(2)}\Big ),  \label{H2t4}
\end{align}
respectively, where $V_{n}\ (n=1,2,3,4)$ are the same hybridization
integrals defined in Eqs. (\ref{hopping}). As shown in the previous
subsection, due to the oxygen atoms displacements, there are more active
orbitals involved. Relevant states in such a four-atom cluster include $%
|P_{j}\rangle$, $|AP_{j}\rangle$ and $p_{\mu,\sigma}^{(i)}$, where $i,j=1,2$%
, $\mu=x,y,z$, $\sigma=\uparrow,\downarrow$.

In the similar way as the three-atom cluster, the two lowest lying states of
the four-atom cluster can be obtained with energy levels,
\begin{equation}  \label{E4atom}
E_{1(2)} = 2(C \mp \left\vert A_{1}\alpha\right\vert ),
\end{equation}
and corresponding eigenvectors $\left\vert 1\right\rangle$ and $\left\vert
2\right\rangle$ given in Appendix \ref{App4}, where the parameters $C$, $A_1$
and $\alpha$ are those given in Eqs. (\ref{CB}), (\ref{A12}) and (\ref%
{alphabeta}).

\section{Electric Polarization in Clusters}

\label{Sec:P-Cluster}

We shall calculate electric polarizations $\vec{P}=\left\langle e\vec{r}%
\right\rangle $ for three-atom M-O-M clusters or four-atom M-O$_2$-M
clusters using the states $\left\vert 1\right\rangle$ and $\left\vert
2\right\rangle$ obtained from cluster models in previous section. The
position of Fermi energy $E_F$ is crucial to determine the expectation value
$\vec{P}$, which is given by
\begin{equation}
\vec{P}=n_{F}(E_1-E_F)\frac{\left\langle 1\right\vert e\vec{r}\left\vert
1\right\rangle}{\left\langle 1|1\right\rangle } +n_{F}(E_2-E_F)\frac{%
\left\langle 2\right\vert e\vec{r}\left\vert 2\right\rangle }{\left\langle
2|2\right\rangle },  \label{polar-super}
\end{equation}
where $n_{F}(E)=\frac{1}{1+e^{\beta E}}$ is the Fermi function. At low
temperature, one may replace $n_{F}(E)$ by $\theta(-E)$. In the case of
single-hole, say, $E_1<E_F<E_2$, only the lowest state $\left\vert
1\right\rangle $ contributes to electric polarization. While in the
double-hole situation, $E_F>E_2$, both state $\left\vert 1\right\rangle $
and $\left\vert 2\right\rangle $ are active \cite{Katsura05}.

According to the expression of $\left\vert 1\right\rangle$ and $\left\vert
2\right\rangle$ given in Appendix \ref{App3} and Appendix \ref{App4}, the
electric polarization $\vec{P}$ can be written in terms of the overlap
dipole matrix elements $I^{\alpha}_{\mu,\nu}(\vec{a})$ given as follows,
\begin{equation}  \label{Eq.defI}
I^{\alpha}_{\mu,\nu}(\vec{a})=e\left\langle d_{\mu}(\vec{r})\right\vert
\alpha \left\vert p_{\nu}(\vec{r}+\vec{a})\right\rangle,
\end{equation}
where $\vec{a}$ is the displacement from a transition metal atom M to one of
its neighboring oxygen atoms O, $\alpha=x,y,z$ is one of the three
components of $\vec{r}$, $\mu=xy,yz,zx$ denotes three $t_{2g}$ orbitals and $%
\nu=x,y,z$ denotes three oxygen $p$-orbitals. We calculate $%
I^{\alpha}_{\mu,\nu}(\vec{a})$ in Appendix \ref{AppI} and study the
configurations of electric polarization in the following subsections.

We find that the electric polarization consists of two parts. The first part
can be written in terms of $\vec{e}_{j}\times\vec{e}_{j+1}$, which is called
``spin-current polarization". The second part depends on the bond bending
angle and distortion configuration, which is called ``lattice-mediated
polarization".

\subsection{Three-atom M-O-M cluster}

For a three-atom M-O-M cluster, both spin-current polarization and
lattice-mediated polarization contribute to the electric polarization.
However, they may play different roles in different situations. We shall
discuss singe-hole and double-hole situations respectively.

For \textit{single-hole situation} when $E_{1}<E_{F}<E_{2}$, only the lowest
state $\left\vert 1\right\rangle $ contributes to electric polarization. The
induced electric polarization $\vec{P}_{j,j+1}$ at each element cluster
connecting $j$ and $j+1$-th transition metal atoms can be calculated with
the help of Eq. (\ref{s-1}). The resulting polarization with three
components reads,
%
\begin{align}
P_{j,j+1}^{x}& =0,  \notag \\
P_{j,j+1}^{y}& =\frac{1}{3}\frac{I_{1}^{y}A_{1}+I_{2}^{y}A_{2}%
}{\Delta \left\vert B_{j,j+1}\right\vert }\left. \left( \vec{e}%
_{j,j+1}\times (\vec{e}_{j}\times \vec{e}_{j+1})\right) \right\vert _{y}
\notag \\
& \ \ \ -\frac{2}{3}\frac{I_{3}^{y}}{%
\Delta }-\frac{2}{3}\frac{I_{2}^{y}A_{1}\left\vert \alpha
_{j,j+1}\right\vert ^{2}+I_{1}^{y}A_{2}\left\vert \beta_{j,j+1}\right\vert ^{2}%
}{\Delta \left\vert B_{j,j+1}\right\vert},
\notag \\
P_{j,j+1}^{z}& =\frac{1}{3}\frac{I_{4}^{z}A_{1}}{\Delta
\left\vert B_{j,j+1}\right\vert }\left. \left( \vec{e}_{j,j+1}\times (\vec{e}%
_{j}\times \vec{e}_{j+1})\right) \right\vert _{z}  \notag \\
& \ \ \ +\frac{1}{3}\frac{I_{4}^{z}A_{2}}{\Delta
\left\vert B_{j,j+1}\right\vert }(\sin \theta _{j}\cos \theta _{j+1}\sin
\phi _{j}\nonumber\\
&\ \ \ \ \ \ \ \ \ \ \ \ \ \ \ \ \ \ \ \ \ \ +\cos \theta _{j}\sin \theta _{j+1}\sin \phi _{j+1}),
\end{align}%
where $B_{j,j+1}=A_{1}\alpha _{j,j+1}+A_{2}\beta _{j,j+1}$, and $\alpha
_{j,j+1}$ and $\beta _{j,j+1}$ are given as follows,
\begin{eqnarray}
\alpha _{j,j+1} &=&\sin \frac{\theta _{j}}{2}\sin \frac{\theta _{j+1}}{2}%
+\cos \frac{\theta _{j}}{2}\cos \frac{\theta _{j+1}}{2}e^{-i(\phi _{j}-\phi
_{j+1})},  \notag \\
\beta _{j,j+1} &=&i(\sin \frac{\theta _{j}}{2}\sin \frac{\theta _{j+1}}{2}%
-\cos \frac{\theta _{j}}{2}\cos \frac{\theta _{j+1}}{2}e^{-i(\phi _{j}-\phi
_{j+1})}),\nonumber\\
&\label{alphabetaj}
\end{eqnarray}%
by replacing $\phi _{1}$ and $\phi _{2}$ in Eq.(\ref{alphabeta}) by $\phi
_{j}$ and $\phi _{j+1}$ respectively. And other parameters are defined as
\begin{eqnarray}
  I_{1}^{y}&=&V_{3}I_{yz,z}^{y}+V_{4}I_{zx,z}^{y},\nonumber\\
  I_{2}^{y}&=&-V_{1}I_{xy,x}^{y}+V_{2}I_{xy,y}^{y}+V_{3}I_{zx,z}^{y}-V_{4}I_{yz,z}^{y}, \nonumber \\
  I_{3}^{y}&=&V_{1}I_{xy,x}^{y}+V_{2}I_{xy,y}^{y}+V_{3}I_{zx,z}^{y}+V_{4}I_{yz,z}^{y},\nonumber\\
  I_{4}^{z}&=&V_{2}I_{yz,y}^{z}+V_{4}I_{xy,z}^{z}-V_{1}I_{yz,x}^{z}.
\end{eqnarray}
Each component in the electric
polarization $\vec{P}_{j,j+1}$ contains two parts, one part can be written
in terms of $\vec{e}_{j}\times \vec{e}_{j+1}$ which has been predicted in
Ref. \cite{Katsura05}, the other part is related to the bond distortion and
depends on the bending angle $\varphi $. We denote the second part subject
to bond distortion as \textquotedblleft lattice-mediated", to distinguish it
from the \textquotedblleft spin-current" polarization in the first part.
Note that, both lattice-mediated contribution and spin-current contribution
are of the order of $V/\Delta$ and comparable to each other when $\varphi $
is large enough. To derive $P^x_{j,j+1}$, we use the condition that the two
M atoms locate mirror-symmetrically about the $yz$ plane across the oxygen atom
and the $\Gamma_7$ manifold in Eq.(\ref{Eq.ab}) which keeps unchanged under the mirror
operation $x\to-x$.

For \textit{double-hole situation} when $E_{F}>E_{2}$, two lowest states $%
\left\vert 1\right\rangle $ and $\left\vert 2\right\rangle $ are involved.
Using Eq.(\ref{polar-super}) and Eqs.(\ref{s-1}) (\ref{s-2}), we can easily
calculate the electric polarization $\vec{P}_{j,j+1}$, which results in
\begin{align}
P_{j,j+1}^{x}& =0,  \notag \\
P_{j,j+1}^{y}& =-\frac{2}{3}\frac{I_{1}^{y}A_{1}+I_{2}^{y}A_{2}%
}{\Delta ^{2}}\left. \left( \vec{e}_{j,j+1}\times (\vec{e}_{j}\times \vec{e}%
_{j+1})\right) \right\vert _{y}  \notag \\
\ \ \ \ \ & \ \ \ -\frac{4}{3}\frac{I_{3}^{y}}{%
\Delta }\Big (1+\frac{C}{\Delta }\Big )+\frac{4}{3}\frac{I_{2}^{y}A_{1}%
\left\vert \alpha _{j,j+1}\right\vert ^{2}+I_{1}^{y}A_{2}\left\vert \beta _{j,j+1}\right\vert ^{2}%
}{\Delta ^{2}},  \notag \\
P_{j,j+1}^{z}& =-\frac{2}{3}\frac{I_{4}^{z}A_{1}}{\Delta ^{2}}%
\left. \left( \vec{e}_{j,j+1}\times (\vec{e}_{j}\times \vec{e}_{j+1})\right)
\right\vert _{z}  \notag \\
& \ \ \ \ -\frac{2}{3}\frac{I_{4}^{z}A_{2}}{\Delta ^{2}}%
(\sin \theta _{j}\cos \theta _{j+1}\sin \phi _{j} \nonumber \\
&\qquad \qquad \qquad +\cos \theta _{j}\sin\theta _{j+1}\sin \phi _{j+1}).
\end{align}%
To derive $\vec{P}_{j,j+1}$, we have made the approximation $\left\langle
1|1\right\rangle ^{-1}\approx 1+\left( C-\left\vert B\right\vert \right)
/\Delta $ and $\left\langle 2|2\right\rangle ^{-1}\approx 1+\left(
C+\left\vert B\right\vert \right) /\Delta $. Similar to the single-hole
situation, the electric polarization can be divided into two parts, the
lattice mediated contribution and the spin-current contribution too.
However, it is different from the single-hole situation that the
spin-current contribution is of the order of $\left( V/\Delta \right) ^{3}$
while the lattice mediated part is of the order of $V/\Delta $.
Because the electric polarization contains
two parts from states $\left\vert 1\right\rangle $ and $\left\vert
2\right\rangle $ respectively which tend to cancel each other \cite%
{Katsura05}. So the
lattice mediated contribution will dominate over spin-current contribution
when $\varphi $ is large enough.

\subsection{Four-atom M-O$_{2}$-M cluster}
\label{subsec:P-four-site}

For a four-atom M-O$_{2}$-M cluster, only the spin-current polarization
contributes to the electric polarization while lattice-mediated polarization
vanishes. Using the results in Appendix \ref{App4}, one can calculate
electric polarization straightforward.

For \textit{single-hole situation}, one obtains that
\begin{eqnarray}
P_{j,j+1}^{x} &  =&0, \nonumber\\
P_{j,j+1}^{y} &  =&\frac{2}{3}\frac{I_{1}^{y}}{\Delta
}\frac{A_{1}}{\left\vert A_{1}\right\vert }\frac{\left.  \Big (\vec{e}%
_{j,j+1}\times(\vec{e}_{j}\times\vec{e}_{j+1})\Big )\right\vert _{y}%
}{\left\vert \alpha_{j,j+1}\right\vert },\nonumber\\
P_{j,j+1}^{z} &  =&\frac{2}{3}\frac{I_{4}^{z}}{\Delta}\frac{A_{1}}{\left\vert A_{1}\right\vert }%
\frac{\left.  \Big (\vec{e}_{j,j+1}\times(\vec{e}_{j}\times\vec{e}%
_{j+1})\Big )\right\vert _{z}}{\left\vert \alpha_{j,j+1}\right\vert },
\end{eqnarray}
where $\alpha_{j,j+1}$ is given in Eq.(\ref{alphabetaj}).

For \textit{double-hole situation}, one obtains that

\begin{eqnarray}
P_{j,j+1}^{x}  &  =&0,\nonumber\\
P_{j,j+1}^{y}  &  =&-\frac{8}{3}\frac{I_{1}^{y}A_{1}%
}{\Delta^{2}}\left.  \Big (\vec{e}_{j,j+1}\times(\vec{e}_{j}\times\vec
{e}_{j+1})\Big )\right\vert _{y},\nonumber\\
P_{j,j+1}^{z}  &  =&-\frac{8}{3}\frac{I_{4}^{z}A_{1}}{\Delta^{2}}\left.  \Big (\vec{e}_{j,j+1}\times
(\vec{e}_{j}\times\vec{e}_{j+1})\Big )\right\vert _{z}.
\end{eqnarray}
One sees that lattice-mediated contribution to the electric polarization is
canceled due to the reflection symmetry between two oxygen atoms in such a
M-O$_{1(2)}$-M cluster.

\section{Chain Models and Electric Orders}

\label{Sec:chain-models}

We go further to investigate different chains formed by three-atom M-O-M
clusters or four-atom M-O$_2$-M clusters and study possible electric orders
in such chains. Since most multiferroics are insulators, relevant electrons
are localized within clusters, we may neglect electron hopping between
clusters and use relevant local states $\left\vert 1\right\rangle$ and $%
\left\vert 2\right\rangle$ obtained from cluster models in previous section
to calculate the electric polarization $\vec{P}$ within each cluster.

\subsection{Uniform bending M-O-M chain}

\label{subsec:up-chain}

\begin{figure}[ptbh]
\centering
\includegraphics[width=8cm]{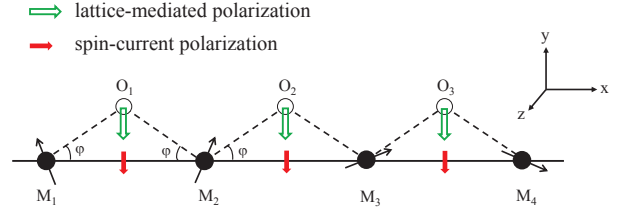}
\caption{(Color online) A sketch of the uniform bending M-O-M chain. All the
bridge oxygen atoms deviate from the transition metal atoms chain in the
same direction and lie within the $xy$-plane. And the spin-spiral plane is $xy$-plane. The spin-current and
lattice-mediated polarization are denoted by red and green arrows,
respectively. }
\label{fig:up-chain}
\end{figure}

Firstly, we consider the uniform bond-distortion chain model shown in Fig.%
\ref{fig:up-chain}. In this case, all the bridge oxygen atoms deviate from
the spin chain formed by transition metal atoms in the same direction and
with the same bond-bending angle $\varphi$. This type of bond-distortion
may be ascribed to the DM interaction relating to spiral spin ordering,
where the spin-helicity vector has the same sign for all pairs of neighbouring spins
in case of transverse-spiral ordering, the DM coupling pushes negative oxygen ions
in the same direction\cite{Chenong}.

It is easy to see that ferroelectric order will form in this kind of chains.
To illustrate it, we assume all the magnetic moment from transition metal
ions lie in $xy$-plane and are spiral ordered in the same plane as shown in Fig.\ref%
{fig:up-chain}. In this case, $P_{j,j+1}^{x}=P_{j,j+1}^{z}=0$, and $%
P_{j,j+1}^{y}$ does not depend on site index $j$.

For single-hole,
\begin{align}
P_{j,j+1}^{y}& =\frac{1}{3}\frac{I_{1}^{y}A_{1}+I_{2}^{y}A_{2}%
}{\Delta}\frac{\sin (\Delta\phi)}{\left\vert A_{1}\cos\frac{\Delta\phi}{2}-A_{2}\sin\frac{\Delta\phi}{2}\right\vert}
\notag \\
& \ \ \ -\frac{2}{3}\frac{I_{3}^{y}}{%
\Delta }-\frac{2}{3}\frac{I_{2}^{y}A_{1}\left\vert \cos\frac{\Delta\phi}{2}\right\vert ^{2}+I_{1}^{y}A_{2}\left\vert \sin\frac{\Delta\phi}{2}\right\vert ^{2}}{\Delta\left\vert A_{1}\cos\frac{\Delta\phi}{2}-A_{2}\sin\frac{\Delta\phi}{2}\right\vert },\notag \\
\end{align}
while for double-hole,
\begin{align}
P_{j,j+1}^{y}  &  =-\frac{2}{3}\frac{I_{1}^{y}A_{1}+I_{2}^{y}A_{2}}{\Delta^{2}}\sin (\Delta\phi)%
-\frac{4}{3}\frac{I_{3}^{y}}{\Delta}\Big (1+\frac{C}{\Delta%
}\Big )\nonumber\\
\ \ \ \ \  &  \ \ \ +\frac{4}{3}\frac{I_{2}^{y}A_{1}\left\vert \cos\frac{\Delta\phi}{2}\right\vert ^{2}
+I_{1}^{y}A_{2}\left\vert \sin\frac{\Delta\phi}{2}\right\vert ^{2}}{\Delta^{2}},%
\end{align}
with $\Delta\phi=\phi_{j}-\phi_{j+1}$.

\subsection{Staggered bending M-O-M chain}

\label{subsec:up-down-chain}

\begin{figure}[ptbh]
\centering
\includegraphics[width=8cm]{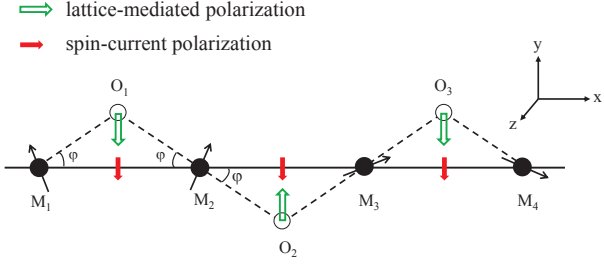}
\caption{(Color online) A sketch of the staggered M-O-M chain. The neighbor
oxygen atoms of a given magnetic metal atom displace toward opposite
directions. The spin-current polarization is uniform in each M-O-M cluster
(red arrows), while lattice-mediated polarization is staggered (green
arrows).}
\label{fig:up-down chain}
\end{figure}

There also exist some materials with staggered bending M-O-M chains as shown
in Fig.\ref{fig:up-down chain}. For instance, the Mn-O-Mn-O-Mn bonding along
$\left\langle 110\right\rangle $ direction forms a zigzag chain due to
alternative rotation and tilt of the MnO$_{6}$ octahedra in some perovskite
rare earth manganese oxides RMnO$_{3}$\cite{Arima}.

In this case, spin-current polarization and lattice-mediated polarization
may form different orders. To illustrate it, we still assume all the
magnetic moment from transition metal ions lie in $xy$-plane and are spiral
ordered as shown in Fig.\ref{fig:up-down chain}. In this case, $%
P_{j,j+1}^{x}=P_{j,j+1}^{z}=0$. $P_{j,j+1}^{y}$ contains two parts,
spin-current polarization $P_{j,j+1}^{sc,y}$ and lattice-mediated
polarization $P_{j,j+1}^{lm,y}$,
\begin{equation}
P_{j,j+1}^{y}=P_{j,j+1}^{sc,y}+P_{j,j+1}^{lm,y}.
\end{equation}
For single-hole,
\begin{align}
P_{j,j+1}^{sc,y}  &  =\frac{1}{3}\frac{I_{1}^{y}A_{1}+I_{2}^{y}A_{2}}{\Delta}\frac{\sin(\Delta\phi)}{\left\vert A_{1}\cos\frac{\Delta\phi}{2}+(-1)^{j}A_{2}\sin\frac{\Delta\phi}{2}
\right\vert },\nonumber\\
P_{j,j+1}^{lm,y}&=(-1)^{j}\frac{2}{3}\frac{I_{3}^{y}}{%
\Delta }  \notag \\
& \ \ \ +(-1)^{j}\frac{2}{3}\frac{I_{2}^{y}A_{1}\left\vert
\cos\frac{\Delta\phi}{2}\right\vert ^{2}+I_{1}^{y}A_{2}\left\vert \sin\frac{\Delta\phi}{2}\right\vert ^{2}}{\Delta\left\vert A_{1}\cos\frac{\Delta\phi}{2}+(-1)^{j}A_{2}\sin\frac{\Delta\phi}{2}\right\vert}.
\end{align}
while for double-hole,
\begin{align}
P_{j,j+1}^{sc,y}  &  =-\frac{2}{3}\frac{I_{1}^{y}A_{1}+I_{2}^{y}A_{2}}{\Delta^{2}}\sin(\Delta\phi),\nonumber\\
P_{j,j+1}^{lm,y}&=(-1)^{j}\frac{4}{3}\frac{I_{3}^{y}}{%
\Delta }\Big (1+\frac{C}{\Delta
}\Big )  \notag \\
& +(-1)^{j+1}\frac{4}{3}\frac{I_{2}^{y}A_{1}\left \vert \cos\frac{\Delta\phi}{2} \right \vert^{2}+I_{1}^{y}A_{2}\left \vert \sin\frac{\Delta\phi}{2}\right \vert^{2}}{\Delta^{2}}.
\end{align}

So that spin-current polarization is uniform and forms ferroelectric order
while lattice-mediated polarization is staggered and forms antiferroelectric
order. The total electric polarization is \textit{``ferrielectric ordered"}.

\subsection{Symmetric M-O$_{2}$-M chain}

\label{subsec:symmetric-chain}

Finally, we consider a symmetric M-O$_{2}$-M chain shown in Fig.\ref%
{fig:symmetric-chain}, which is made of four-atom M-O$_{2}$-M clusters. Such
symmetric M-O$_{2}$-M chains exist in two prominent examples of multiferroic
compound, namely, LiCu$_{2}$O$_{2}$ \cite{LCuO07,LCuO08} and LiCuVO$_{4}$
\cite{LCuVO07}, which crystallize as one-dimensional chains of the
edge-sharing CuO$_{4}$ plaquettes.

\begin{figure}[ptbh]
\centering
\includegraphics[width=8cm]{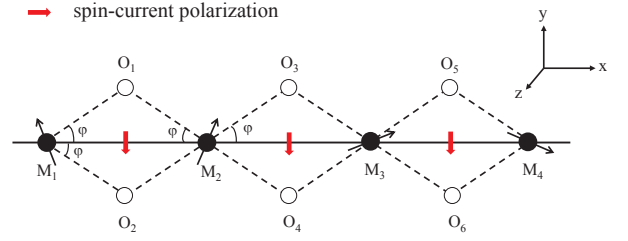}
\caption{(Color online) A schematic edge-shared MO$_{2}$ chain. The bridge
oxygen atoms are symmetrically located beside the two sides of the magnetic
metal atoms chain. In this case, the lattice-mediated contribution of the
electric polarization is exactly canceled. The nonvanish part of the
polarization arise from the spin current, indicated by red arrows. }
\label{fig:symmetric-chain}
\end{figure}

In this kind of chains, lattice-mediated polarization vanishes and
spin-current polarization may form ferroelectric order. Assuming all the
magnetic moment from transition metal ions lie in $xy$-plane and are spiral
ordered as shown in Fig.\ref{fig:symmetric-chain}, we obtain $%
P_{j,j+1}^{x}=P_{j,j+1}^{z}=0$ and
\begin{align}
P_{j,j+1}^{y} &  =\frac{2}{3}\frac{I_{1}^{y}A_{1}}%
{\Delta\left\vert A_{1}\right\vert }\frac{\sin(\Delta\phi)}%
{\left \vert \cos \frac{\Delta\phi}{2}\right \vert},%
\end{align}
in single-hole situation and
\begin{align}
P_{j,j+1}^{y}  &  =-\frac{8}{3}\frac{I_{1}^{y}A_{1}%
}{\Delta^{2}}\sin (\Delta\phi),
\end{align}
in double-hole situation.

\section{Discussions}

\label{Sec:discussion}

In this section, we shall discuss the relation between magnetism-driven electric polarization and lattice-driven electric polarization,
and then apply our theory to multiferroic copper oxides.

\begin{figure}[hbtp]
  \centering
  \includegraphics[width=8cm]{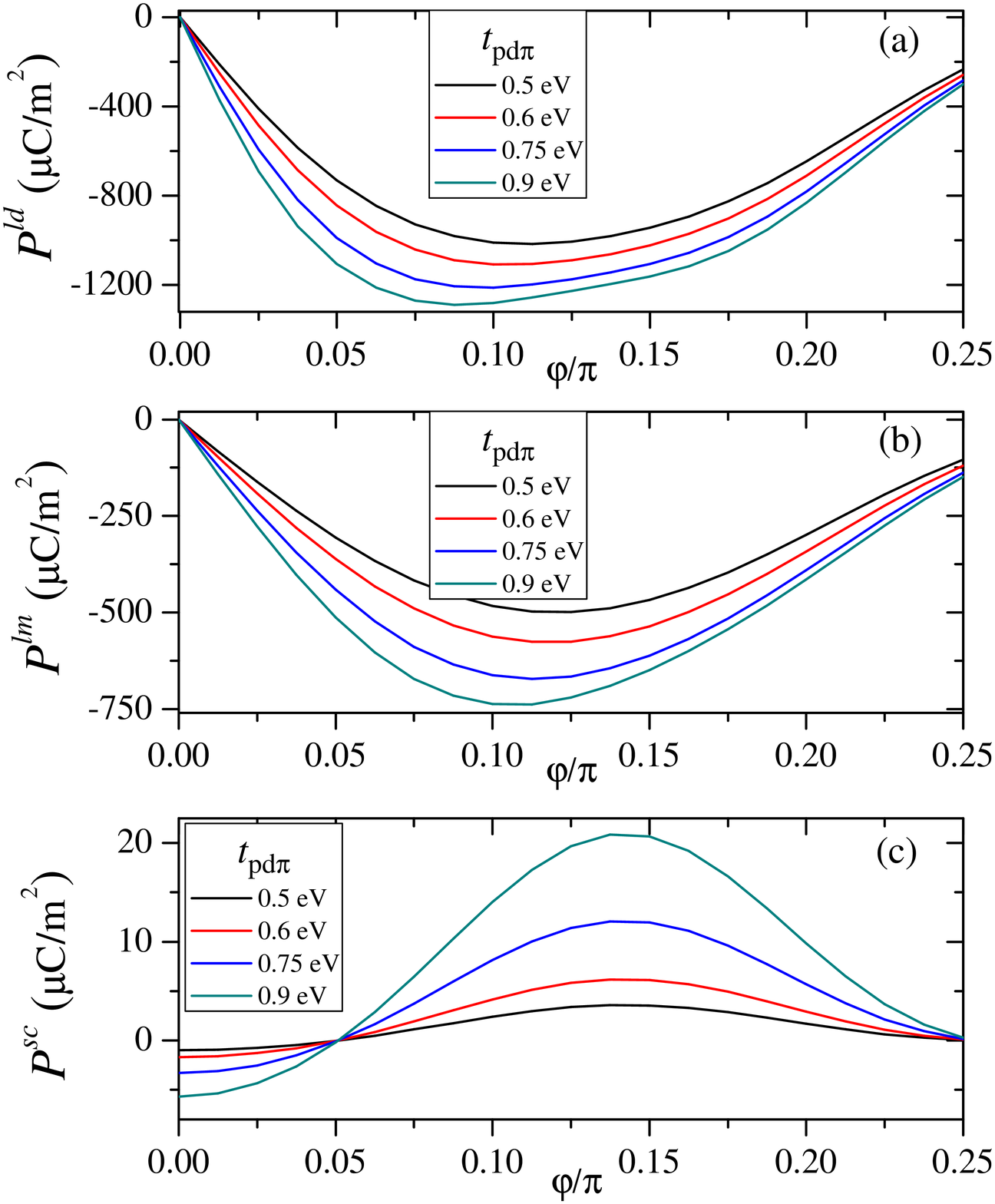}
  \caption{(Color online) The $\varphi$-dependence of electric polarizations in uniform bending M-O-M chains (see Fig.\ref{fig:up-chain}).
  (a) Lattice-driven electric polarization. (b) Lattice-mediated polarization. (c) Spin-current polarization.}
  \label{Fig:UniformChainP}
\end{figure}

{\it Lattice-driven electric polarization:}
As shown in Fig.\ref{fig:up-chain}, the negatively charged oxygens coherently displace
away from the magnetic chain formed by M ions. This shift will generate a net
\emph{lattice-driven} electric polarization $P^{ld}$ directing from the oxygen ions
to the M ions chain. The polarization $P^{ld}$ can be evaluated as long as the $p$-$d$
hybridizations  $t_{pd\sigma}$ and $t_{pd\pi}$ and the geometry of the lattice are specified.
We shall compare this lattice-driven electric polarization ($P^{ld}$) with
the two parts of magnetism-driven electric polarization studied in this paper, say,
lattice-mediated polarization ($P^{lm}$) and spin-current polarization ($P^{sc}$).
To do this, we restrict ourselves to the double-hole situation and choose the following
parameters, $t_{pd\sigma}/t_{pd\pi}=-2$,\cite{tpd-tpi} $\Delta=2$eV, $\sin(\Delta\phi)=\sin(0.28\pi)$\cite{Jia2007,Hu07,Hu10}
and the lattice constant $a=5$\AA. The results are shown in Fig.\ref{Fig:UniformChainP}
where electric polarizations are plotted as functions of bond-bending angle $\varphi$.
For a typical $\varphi$, $P^{ld}\sim10^{2}\mu$C/m$^{2}$, while $P^{lm}\sim10^{2}\mu$C/m$^{2}$
and $P^{sc}\sim10\mu$C/m$^{2}$.
The lattice-mediated polarization keeps in parallel with the lattice-driven polarization
for any bond-bending angle $\varphi$ while the spin current polarization may varies with
$\varphi$ and be in parallel or anti-parallel to the lattice-driven polarization.
Moreover, the spin current polarization depends on spin configuration and can be neither parallel
nor anti-parallel to lattice-driven polarization. Upon these observations,
we conclude that lattice-driven polarization is cooperative with lattice-mediated polarization
while spin current polarization may be either competitive or cooperative,
depending on the spin configuration and the bond-bending angle.

Note that lattice-driven polarization $P^{ld}$ in Fig.\ref{Fig:UniformChainP}(a)
is overestimated since we chose the maximum electric polarization in all the possible
valences for magnetic transition metal and oxygen ions.
It is also worth noting that altough $P^{ld}$ and $P^{lm}$ dominate over $P^{sc}$ at
finite $\varphi$ in uniform bending chains as shown in Fig.\ref{Fig:UniformChainP},
$P^{sc}$ will be dominating at small $\varphi$ or in symmetric M-O$_{2}$-M chains.

{\it Multiferroic copper oxides:}
Then we shall apply our theory to several copper oxides, including LiCuVO$_{4}$, LiCu$_{2}$O$_{2}$ and NaCu$_{2}$O$_{2}$,
where magnetic copper ions Cu$^{2+}$ and oxygen ions O$^{2-}$ form symmetric M-O$_{2}$-M chain as shown in Fig.\ref{fig:symmetric-chain}.

There are two prominent compounds which can be described symmetric M-O$_{2}$-M chains, LiCu$_{2}$O$_{2}$
and LiCuVO$_{4}$. Both of them are characterized by edge-sharing CuO$_{4}$ plaquettes,
forming spiral spin orders and emerging electric polarization at low temperature.
The electric polarization in these two compounds is relatively weak comparing with manganites, $\sim10\mu$C/m$^{2}$,
and shows strong anisotropy.
There is another interesting material NaCu$_{2}$O$_{2}$, which is isostructural to LiCu$_{2}$O$_{2}$ while Li is substituted by Na.
Although NaCu$_{2}$O$_{2}$ also exhibits helical magnetic order at low temperature\cite{NaCuO-05,NaCuO-06,NaCuO-08},
no electric polarization has been observed in it within the experimental sensitivity ($<$$0.3\mu$C/m$^{2}$)\cite{NaCuO-10-1,NaCuO-10-2}.
We shall apply our theory to these prototypical compounds and address the issue of ``missing multiferroicity'' in NaCu$_{2}$O$_{2}$
in the rest of this section.

As mentioned in Section \ref{subsec:P-four-site}, the lattice-mediated contribution is exactly cancelled
in a symmetric M-O$_{2}$-M cluster, and the electric polarization comes from the spin current only.
In contrast to widely used isotropic form of Eq.(\ref{KNB}),
the spin current form of electric polarization in the symmetric M-O$_{2}$-M cluster is anisotropic in general,
which is in accord with the experimental findings\cite{LCuVO08-Yasui,LCuVO08-schrettle,LCuO07,LCuO08}
and density functional calculations\cite{Xiang07Cuprate}. The magnitude of the electric polarization
depends on the bond-bending angle $\varphi$ and vanishes at $\varphi=\pi/4$. Note that the spin current form
will recover isotropic in the absence of bond distortion, say, $\varphi=0$.

\begin{figure}[hbtp]
  \centering
  \includegraphics[width=9cm]{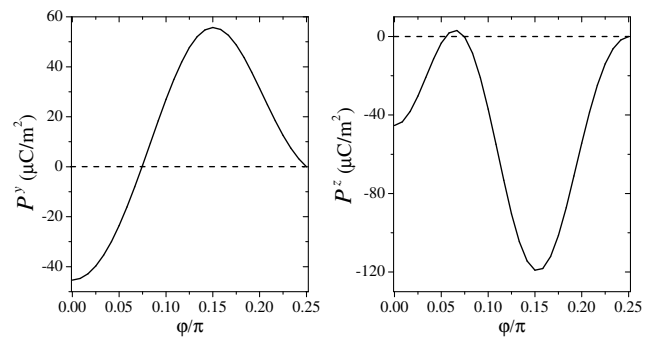}
  \caption{ The $\varphi$-dependence of $y$- and $z$- components of electric polarization
  in the presence of $xy$- and $zx$-plane spin spiral order respectively
in symmetric M-O$_{2}$-M chains (see Fig.\ref{fig:symmetric-chain}).
A typical set of parameters are choosen for Cu$^{2+}$ oxides\cite{Pickett99}, $t_{pd\pi}=0.75$eV, $t_{pd\sigma}=-1.85$eV,
  $\Delta=3$eV. The Clementi-Raimondi effective charges $Z_{\mathrm{O}}^{\mathrm{eff}}=4.45$ and $Z_{\mathrm{Cu}}^{\mathrm{eff}}=13.2$
are used for O$^{2-}$ and Cu$^{2+}$ ions respectively\cite{CR-1} and
R$_{\mathrm{Cu-O}}=1.445$\AA \ (i.e., LiCuVO$_{4}$\cite{Whangbo11}).}
  \label{Fig:SymmetryChainP}
\end{figure}

In Fig.\ref{Fig:SymmetryChainP}, we plot $y$- and $z$- components of electric polarization
as functions of bond-bending angle $\varphi$ in the presence of $xy$- and $zx$-plane spin spiral order respectively.
One sees that the polarization is sensitive to the bond bending angle $\varphi$ and is anisotropic.
In realistic materials, the bond angles of Cu-O-Cu in LiCuVO$_{4}$, LiCu$_{2}$O$_{2}$, and NaCu$_{2}$O$_{2}$
are $95^{\circ}$ \cite{LCuVO08-Yasui}, $94^{\circ}$ \cite{LCuO08}, and $92.9^{\circ}$ \cite{NaCuO-05,NaCuO-06} respectively,
corresponding to bond bending angles $\varphi= 42.5^{\circ}, 43^{\circ}$, and $43.6^{\circ}$,
which are all close to $\varphi=\pi/4$.
As shown in the Fig. \ref{Fig:SymmetryChainP}, the electric polarization decreases and ultimately vanishes
when $\varphi$ approaches $\pi/4$.
Indeed, experiments clearly indicated that the electric polarization in LiCuVO$_{4}$ is
stronger than that of LiCu$_{2}$O$_{2}$. Moreover, the bond-bending angle $\varphi$
in NaCu$_{2}$O$_{2}$ is closer to $\pi/4$ than that of LiCuVO$_{4}$ and LiCu$_{2}$O$_{2}$,
resulting in weaker electric polarization which is difficult to be observed.

\begin{table}[hbtp]
\caption{\label{table:PolarCompare}
The $a$- and $c$- components of electric polarizations in LiCuVO$_{4}$, LiCu$_{2}$O$_{2}$ and NaCu$_{2}$O$_{2}$
in the presence of $ab$- and $bc$-plane spin spiral orders.
The values in parentheses correspond to experimental data.
}
\begin{ruledtabular}
\begin{tabular}{l|c|cc}
                      &                       & \multicolumn{2}{c}{Polarization ($\mu$C/m$^2$)} \\
Materials             &$\varphi$              & $P_{a}$($ab$)         & $P_{c}$($bc$) \\
\hline
LiCuVO$_{4}$\cite{LCuVO08-Yasui,LCuVO08-schrettle}  & 42.5$^{\circ}$       & 5.8 (43)                 & 4.3 (10)  \\
LiCu$_{2}$O$_{2}$\cite{LCuO07,LCuO08}               & 43$^{\circ}$         & 4.4 (8)                  & 3.8 (4)   \\
NaCu$_{2}$O$_{2}$\cite{NaCuO-10-2}                  & 43.6$^{\circ}$       & 2.1 (?\footnotemark[1])  & 0.8 (0)   \\

\end{tabular}
\footnotemark[1]{Currently no experimental data is available. }
\end{ruledtabular}
\end{table}

For comparison, we also list calculated electric polarizations for these three compounds in
Table \ref{table:PolarCompare} using their concrete parameters.
For example, the pitch angles for LiCuVO$_{4}$, LiCu$_{2}$O$_{2}$ and NaCu$_{2}$O$_{2}$
are 84.2$^\circ$\cite{LCuVO08-Yasui,LCuVO08-schrettle}, 62.6$^\circ$\cite{LCuO08},
and 81.7$^\circ$\cite{NaCuO-10-2} respectively. Corresponding R$_{\mathrm{Cu-O}}$
(R$_{\mathrm{Cu-O}}$ is the distance of Cu-O bond along the Cu-Cu chain)
are 1.445\AA \cite{Whangbo11}, 1.43\AA \cite{LCuO08}, and 1.465\AA \cite{NaCuO-10-2}
respectively. Other parameters are the same as those in Fig.\ref{Fig:SymmetryChainP}.
The values in parentheses in Table \ref{table:PolarCompare} are experimental data.
The decreasing electric polarization with $\varphi$ approaching $\pi/4$ is in agreement with
experimental observation.
For NaCu$_{2}$O$_{2}$ compound in the presence of $bc$-plane spin spiral order\cite{NaCuO-05,NaCuO-08},
we find that $P_{c}$ is extremely weak and hard to detect\cite{NaCuO-10-2}. To our knowledge,
there is no experimental report on electric polarization in the presence of $ab$-plane spin spiral order,
which may be tunable under external field. Our theory predict a finite electric polarization
($\sim 2\mu$C/m$^{2}$) oriented along the $a$-axis in the presence of $ab$-plane spin spiral order.
This prediction can be tested in future experiment for NaCu$_{2}$O$_{2}$.
Thus, the effect of bond-distortion on the magnetic-driven polarization provides a clue
to clarify the anisotropy of the electric polarization observed in LiCuVO$_{4}$ and LiCu$_{2}$O$_{2}$
and to elucidate the puzzle of non-multiferroicity in NaCu$_{2}$O$_{2}$.

\section{Conclusion}

\label{Sec:conclusion}

In this paper, we study how the distortion of M-O-M bonds affect the lowest
lying electronic states thereby the electric polarization in multiferroic
magnets. We calculate relevant low lying electronic states and electric
polarization for three-atom M-O-M clusters and four-atom M-O$_{2}$-M
clusters respectively. It turns out that the electric polarization contains
two parts, spin-current polarization which can be written in terms of $\vec{e%
}_{i,j}\times(\vec{e}_{i}\times\vec{e}_{j})$ and lattice-mediated polarization
which exists only when the M-O-M bond angle deviates away from $180^{\circ}$.

For three-atom M-O-M clusters, both the spin-current polarization and the
lattice-mediated polarization are of the order of $(V/\Delta)$ in
single-hole situation. However, the spin-current polarization is of the
order of $(V/\Delta)^3$ while the lattice-mediated polarization is of the
order of $(V/\Delta)$ in double-hole situation. Thus, the electric
polarization in double-hole situation can be much larger than that estimated
by spin-current mechanism\cite{Katsura05} when M-O-M bond distortion is
present. For four-atom M-O$_{2}$-M clusters, lattice-mediated polarization
vanishes due to symmetry and there exists only spin-current polarization
which is of the order of $(V/\Delta)$ in single-hole situation and of the
order of $(V/\Delta)^3$ in double-hole situation.

Then we study three kinds of chain models made of different clusters where
magnetic moments are spiral ordered. Firstly, we study uniform
bond-distortion chain models where all the bridge oxygen atoms deviate from
the spin chain formed by transition metal atoms in the same direction and
with the same bond-bending angle. Ferroelectric order may exist in such
chains. Secondly, we study staggered bending M-O-M chains in which the
bridge oxygen atoms displace on opposite directions between two adjacent
clusters. Ferrielectric order may appear instead of ferroelectric order in
this kind of chains. Thirdly, we study symmetric M-O$_{2}$-M chains which
are made of four-atom M-O$_{2}$-M clusters and the two oxygen atoms deviate
symmetrically from the M-M bond within a M-O$_{2}$-M cluster. The electric
polarization may be ferroelectric ordered in such symmetric chains.

We also discuss the relation between magnetic-driven electric polarization
and lattice-driven electric polarization in the presence of bond distortion
in details.

Finally, we apply our theory to multiferroics copper oxides and find the results agree
with experimental observations. The issue of ``missing multiferroicity'' in
NaCu$_{2}$O$_{2}$ compound is clarified and more experimental predictions are made.

Summarizing, we show that the M-O-M bond distortion may affect electric
polarization in multiferroics significantly and the possibilities for
ferroelectric and ferrielectric ordering are discussed. The application to
multiferroics copper oxides is successful.

\section*{Acknowledgements}

This work is supported by National Basic Research Program of China(973
Program, No.2011CBA00103), NSFC(Grant No.11074216, 10874149 and 11074218),
RFDP(No.20100101120002), PCSIRT(No.IRT0754), Doctoral Fund of Ministry of Education of China,
and the Fundamental Research Funds for the Central Universities in China.

\appendix

\section{Coefficients $A_{(j)}^{i\sigma}$ and $B_{(j)}^{i\sigma}$ in Eqs.(\ref{Eq.PAPt2g})}

\label{App1}

In this appendix we shall derive coefficients $A_{(j)}^{i\sigma}$ and $%
B_{(j)}^{i\sigma}$ in Eqs.(\ref{Eq.PA}) and (\ref{Eq.APB}). For the parallel
state $\left\vert P_{j}\right\rangle$ and corresponding coefficients $%
A_{(j)}^{i\sigma}$, we combine Eqs.(\ref{Eq.P}) and (\ref{Eq.PA}) to have
\begin{equation}
\sin\frac{\theta_{j}}{2}\left\vert a\right\rangle +e^{i\phi_{j}}\cos\frac{%
\theta_{j}}{2}\left\vert b\right\rangle
=\sum_{i\sigma}A_{(j)}^{i\sigma}\left\vert d_{i\sigma}^{(j)}\right\rangle.
\end{equation}
Using the relation Eq.(\ref{Eq.ab}), we obtain $A_{(j)}^{i\sigma}$ as
follows,
\begin{subequations}
\label{Eq.Aj}
\begin{eqnarray}
A_{(j)}^{xy,\uparrow} & = & \frac{1}{\sqrt{3}}\sin\frac{\theta_{j}}{2}, \\
A_{(j)}^{xy,\downarrow} & = & \frac{1}{\sqrt{3}}e^{i\phi_{j}}\cos\frac{%
\theta_{j}}{2}, \\
A_{(j)}^{yz,\uparrow} & = & -\frac{1}{\sqrt{3}}e^{i\phi_{j}}\cos\frac{%
\theta_{j}}{2}, \\
A_{(j)}^{yz,\downarrow} & = & \frac{1}{\sqrt{3}}\sin\frac{\theta_{j}}{2}, \\
A_{(j)}^{zx,\uparrow} & = & \frac{1}{\sqrt{3}}ie^{i\phi_{j}}\cos\frac{%
\theta_{j}}{2}, \\
A_{(j)}^{zx,\downarrow} & = & \frac{1}{\sqrt{3}}i\sin\frac{\theta_{j}}{2}.
\end{eqnarray}

Similarly, we obtain coefficients $B_{(j)}^{i\sigma}$ for the anti-parallel
state $\left\vert AP_{j}\right\rangle$ as follows,
\end{subequations}
\begin{subequations}
\label{Eq.Bj}
\begin{eqnarray}
B_{(j)}^{xy,\uparrow} & = & \frac{1}{\sqrt{3}}\cos\frac{\theta_{j}}{2}, \\
B_{(j)}^{xy,\downarrow} & = & -\frac{1}{\sqrt{3}}e^{i\phi_{j}}\sin\frac{%
\theta_{j}}{2}, \\
B_{(j)}^{yz,\uparrow} & = & \frac{1}{\sqrt{3}}e^{i\phi_{j}}\sin\frac{%
\theta_{j}}{2}, \\
B_{(j)}^{yz,\downarrow} & = & \frac{1}{\sqrt{3}}\cos\frac{\theta_{j}}{2}, \\
B_{(j)}^{zx,\uparrow} & = & -\frac{1}{\sqrt{3}}ie^{i\phi_{j}}\sin\frac{%
\theta_{j}}{2}, \\
B_{(j)}^{zx,\downarrow} & = & \frac{1}{\sqrt{3}}i\cos\frac{\theta_{j}}{2}.
\end{eqnarray}
\end{subequations}
\section{Lowest lying eigenstates for three-atom M-O-M cluster}\label{App3}

We calculate the eigenvectors for two lowest lying eigenstates $\left\vert
1\right\rangle$ and $\left\vert 2\right\rangle$ in a three-atom M-O-M
cluster in this appendix. The Hilbert space contains states $|P_{j}\rangle$,
$(j=1,2)$ and $p_{\mu,\sigma}$ $(\mu=x,y,z,\ \sigma=\uparrow,\downarrow)$.
Assuming $V\ll\Delta$ and treating $H_{t}$ as a perturbation to $H_{\Delta}$%
, we obtain the two eigenvectors $\left\vert 1\right\rangle$ and $\left\vert
2\right\rangle$ up to the second order perturbation,
\begin{widetext}%
\begin{align}
\left\vert 1\right\rangle  &  =-\frac{B}{\sqrt{2}|B|}\bigg [\left\vert
P_{1}\right\rangle -\frac{1}{\Delta}\sum_{\sigma}\left(  A_{(1)}^{xy,\sigma
}\left(  V_{1}\left\vert p_{x,\sigma}\right\rangle +V_{2}\left\vert
p_{y,\sigma}\right\rangle \right)  +(V_{3}A_{(1)}^{zx,\sigma}+V_{4}%
A_{(1)}^{yz,\sigma})\left\vert p_{z,\sigma}\right\rangle \right)
\bigg ]\nonumber\\
&  \ \ \ \ +\frac{1}{\sqrt{2}}\bigg [\left\vert P_{2}\right\rangle -\frac
{1}{\Delta}\sum_{\sigma}\left(  A_{(2)}^{xy,\sigma}\left(  V_{1}\left\vert
p_{x,\sigma}\right\rangle -V_{2}\left\vert p_{y,\sigma}\right\rangle \right)
+(-V_{3}A_{(2)}^{zx,\sigma}+V_{4}A_{(2)}^{yz,\sigma})\left\vert p_{z,\sigma
}\right\rangle \right)  \bigg ],\label{s-1}%
\end{align}
with $E_{1}=2(C-\left\vert B\right\vert $), and
\begin{align}
\left\vert 2\right\rangle  &  =\frac{B}{\sqrt{2}|B|}\bigg [\left\vert
P_{1}\right\rangle -\frac{1}{\Delta}\sum_{\sigma}\left(  A_{(1)}^{xy,\sigma
}\left(  V_{1}\left\vert p_{x,\sigma}\right\rangle +V_{2}\left\vert
p_{y,\sigma}\right\rangle \right)  +(V_{3}A_{(1)}^{zx,\sigma}+V_{4}%
A_{(1)}^{yz,\sigma})\left\vert p_{z,\sigma}\right\rangle \right)
\bigg ]\nonumber\\
&  \ \ \ \ +\frac{1}{\sqrt{2}}\bigg [\left\vert P_{2}\right\rangle -\frac
{1}{\Delta}\sum_{\sigma}\left(  A_{(2)}^{xy,\sigma}\left(  V_{1}\left\vert
p_{x,\sigma}\right\rangle -V_{2}\left\vert p_{y,\sigma}\right\rangle \right)
+(-V_{3}A_{(2)}^{zx,\sigma}+V_{4}A_{(2)}^{yz,\sigma})\left\vert p_{z,\sigma
}\right\rangle \right)  \bigg ],\label{s-2}%
\end{align}
\end{widetext}
with $E_{2}=2(C+\left\vert B\right\vert $), where parameters $C$ and $B$ are
given in Eqs. (\ref{CB}), (\ref{A12}) and (\ref{alphabeta}) in the main
text, coefficients $A_{(j)}^{i\sigma}$ are given in Eqs.(\ref{Eq.Aj}). It is
easy to verify that the states $\left\vert1\right\rangle $ and $\left\vert
2\right\rangle $ are orthogonal to each other and their normalization
factors are $\left\langle 1|1\right\rangle =1-\left( C-\left\vert
B\right\vert \right) /\Delta$ and $\left\langle 2|2\right\rangle =1-\left(
C+\left\vert B\right\vert \right) /\Delta$ respectively.

\section{Lowest lying eigenstates for four-atom M-O$_2$-M cluster}\label{App4}

In this appendix, we calculate the eigenvectors for two lowest lying
eigenstates $\left\vert 1\right\rangle$ and $\left\vert 2\right\rangle$ in a
four-atom M-O$_2$-M cluster. In this situation, the Hilbert space contains
states $|P_{j}\rangle$ $(j=1,2)$ and $p^{(i)}_{\mu,\sigma}$ $(i=1,2,\
\mu=x,y,z,\ \sigma=\uparrow,\downarrow)$. Up to the second order of $\frac{V%
}{\Delta}$, we obtain the two eigenvectors $\left\vert 1\right\rangle $ and $%
\left\vert 2\right\rangle$ as follows,
\begin{widetext}%
\begin{align}
\left\vert 1\right\rangle  &  =-\frac{A_{1}\alpha}{\sqrt{2}\left\vert
A_{1}\alpha\right\vert }\bigg [\left\vert P_{1}\right\rangle -\frac{1}{\Delta
}\sum_{\sigma}\Big (A_{(1)}^{xy,\sigma}\left(  V_{1}\left\vert p_{x,\sigma
}^{(1)}\right\rangle +V_{2}\left\vert p_{y,\sigma}^{(1)}\right\rangle \right)
+\left(  V_{3}A_{(1)}^{zx,\sigma}+V_{4}A_{(1)}^{yz,\sigma}\right)  \left\vert
p_{z,\sigma}^{(1)}\right\rangle \nonumber\\
&  \ \ \ \ \ \ \ \ \ \ \ \ \ \ \ \ \ \ \ \ \ \ \ \ \ \ \ \ \ \ -A_{(1)}%
^{xy,\sigma}\left(  V_{1}\left\vert p_{x,\sigma}^{(2)}\right\rangle
-V_{2}\left\vert p_{y,\sigma}^{(2)}\right\rangle \right)  +\left(
V_{3}A_{(1)}^{zx,\sigma}-V_{4}A_{(1)}^{yz,\sigma}\right)  \left\vert
p_{z,\sigma}^{(2)}\right\rangle \Big )\bigg ]\nonumber\\
&  \ \ \ \ +\frac{1}{\sqrt{2}}\bigg [\left\vert P_{2}\right\rangle -\frac
{1}{\Delta}\sum_{\sigma}\Big (A_{(2)}^{xy,\sigma}\left(  V_{1}\left\vert
p_{x,\sigma}^{(1)}\right\rangle -V_{2}\left\vert p_{y,\sigma}^{(1)}%
\right\rangle \right)  +\left(  -V_{3}A_{(2)}^{zx,\sigma}+V_{4}A_{(2)}%
^{yz,\sigma}\right)  \left\vert p_{z,\sigma}^{(1)}\right\rangle \nonumber\\
&  \ \ \ \ \ \ \ \ \ \ \ \ \ \ \ \ \ \ \ \ \ \ \ \ \ \ \ \ \ \ -A_{(2)}%
^{xy,\sigma}\left(  V_{1}\left\vert p_{x,\sigma}^{(2)}\right\rangle
+V_{2}\left\vert p_{y,\sigma}^{(2)}\right\rangle \right)  +\left(
-V_{3}A_{(2)}^{zx,\sigma}-V_{4}A_{(2)}^{yz,\sigma}\right)  \left\vert
p_{z,\sigma}^{(2)}\right\rangle \Big )\bigg ],
\end{align}
with $E_{1}=2(C-\left\vert A_{1}\alpha\right\vert )$, and
\begin{align}
\left\vert 2\right\rangle  &  =\frac{A_{1}\alpha}{\sqrt{2}\left\vert
A_{1}\alpha\right\vert }\bigg [\left\vert P_{1}\right\rangle -\frac{1}{\Delta
}\sum_{\sigma}\Big (A_{(1)}^{xy,\sigma}\left(  V_{1}\left\vert p_{x,\sigma
}^{(1)}\right\rangle +V_{2}\left\vert p_{y,\sigma}^{(1)}\right\rangle \right)
+\left(  V_{3}A_{(1)}^{zx,\sigma}+V_{4}A_{(1)}^{yz,\sigma}\right)  \left\vert
p_{z,\sigma}^{(1)}\right\rangle \nonumber\\
&  \ \ \ \ \ \ \ \ \ \ \ \ \ \ \ \ \ \ \ \ \ \ \ \ \ \ \ \ \ -A_{(1)}%
^{xy,\sigma}\left(  V_{1}\left\vert p_{x,\sigma}^{(2)}\right\rangle
-V_{2}\left\vert p_{y,\sigma}^{(2)}\right\rangle \right)  +\left(
V_{3}A_{(1)}^{zx,\sigma}-V_{4}A_{(1)}^{yz,\sigma}\right)  \left\vert
p_{z,\sigma}^{(2)}\right\rangle \Big )\bigg ]\nonumber\\
&  \ \ \ \ +\frac{1}{\sqrt{2}}\bigg [\left\vert P_{2}\right\rangle -\frac
{1}{\Delta}\sum_{\sigma}\Big (A_{(2)}^{xy,\sigma}\left(  V_{1}\left\vert
p_{x,\sigma}^{(1)}\right\rangle -V_{2}\left\vert p_{y,\sigma}^{(1)}%
\right\rangle \right)  +\left(  -V_{3}A_{(2)}^{zx,\sigma}+V_{4}A_{(2)}%
^{yz,\sigma}\right)  \left\vert p_{z,\sigma}^{(1)}\right\rangle \nonumber\\
&  \ \ \ \ \ \ \ \ \ \ \ \ \ \ \ \ \ \ \ \ \ \ \ \ \ \ \ \ \ \ -A_{(2)}%
^{xy,\sigma}\left(  V_{1}\left\vert p_{x,\sigma}^{(2)}\right\rangle
+V_{2}\left\vert p_{y,\sigma}^{(2)}\right\rangle \right)  +\left(
-V_{3}A_{(2)}^{zx,\sigma}-V_{4}A_{(2)}^{yz,\sigma}\right)  \left\vert
p_{z,\sigma}^{(2)}\right\rangle \Big )\bigg ],
\end{align}
\end{widetext}with $E_{2}=2(C+\left\vert A_{1}\alpha\right\vert )$, where
parameters $C$, $A_1$ and $\alpha$ are given in Eqs. (\ref{CB}), (\ref{A12})
and (\ref{alphabeta}) in the main text, coefficients $A_{(j)}^{i\sigma}$ are
given in Eqs.(\ref{Eq.Aj}). The normalization factors are $\left\langle
1|1\right\rangle =1-2\left( C-\left\vert A_{1}\alpha\right\vert \right)
/\Delta$ and $\left\langle 2|2\right\rangle =1-2\left( C+\left\vert
A_{1}\alpha\right\vert \right)/\Delta$.

\section{Overlap dipole matrix elements $I^{\alpha}_{\mu,\nu}(\vec{a})$}\label{AppI}

From the definition of the overlap dipole matrix elements $%
I^{\alpha}_{\mu,\nu}(\vec{a})$ in Eq.(\ref{Eq.defI}), and using the mirror
symmetry about the $xy$-plane ($z\to -z$), we find that the following matrix
elements vanish,
\begin{eqnarray*}
&&I^{x}_{xy,z}=I^{x}_{yz,x}=I^{x}_{yz,y}=I^{x}_{zx,x}=I^{x}_{zx,y}=0, \\
&&I^{y}_{xy,z}=I^{y}_{yz,x}=I^{y}_{yz,y}=I^{y}_{zx,x}=I^{y}_{zx,y}=0, \\
&&I^{z}_{xy,x}=I^{z}_{xy,y}=I^{z}_{yz,z}=I^{z}_{zx,z}=0.
\end{eqnarray*}

Then we study symmetry relations for nonvanishing $I^{\alpha}_{\mu,\nu}(\vec{a%
})$ with two vectors $\vec{a}$ and $\vec{b}$, where $\vec{a}$ and $\vec{b}$
lie within the $xy$-plane and are linked by symmetry $P_x(x\to -x)$ or $%
P_y(y\to -y)$. When $\vec{b}=P_{x}\vec{a}$ or $\vec{b}=P_{y}\vec{a}$, we
have
\begin{eqnarray*}
I^{x}_{xy,y}(\vec{a})=I^{x}_{xy,y}(\vec{b})&,&I^{x}_{xy,x}(\vec{a}%
)=-I^{x}_{xy,x}(\vec{b}), \\
I^{x}_{yz,z}(\vec{a})=-I^{x}_{yz,z}(\vec{b})&,&I^{x}_{zx,z}(\vec{a}%
)=I^{x}_{zx,z}(\vec{b}), \\
I^{y}_{xy,x}(\vec{a})=I^{y}_{xy,x}(\vec{b})&,&I^{y}_{xy,y}(\vec{a}%
)=-I^{y}_{xy,y}(\vec{b}), \\
I^{y}_{yz,z}(\vec{a})=I^{y}_{yz,z}(\vec{b})&,&I^{y}_{zx,z}(\vec{a}%
)=-I^{y}_{zx,z}(\vec{b}), \\
I^{z}_{xy,z}(\vec{a})=-I^{z}_{xy,z}(\vec{b})&,&I^{z}_{yz,x}(\vec{a}%
)=-I^{z}_{yz,x}(\vec{b}), \\
I^{z}_{yz,y}(\vec{a})=I^{z}_{yz,y}(\vec{b})&,&I^{z}_{zx,x}(\vec{a}%
)=I^{z}_{zx,x}(\vec{b}), \\
I^{z}_{zx,y}(\vec{a})=-I^{z}_{zx,y}(\vec{b})&,&
\end{eqnarray*}
for nonvanishing $I^{\alpha}_{\mu,\nu}(\vec{a})$.

\begin{figure}[hptb]
\begin{center}
\includegraphics[width=7cm]{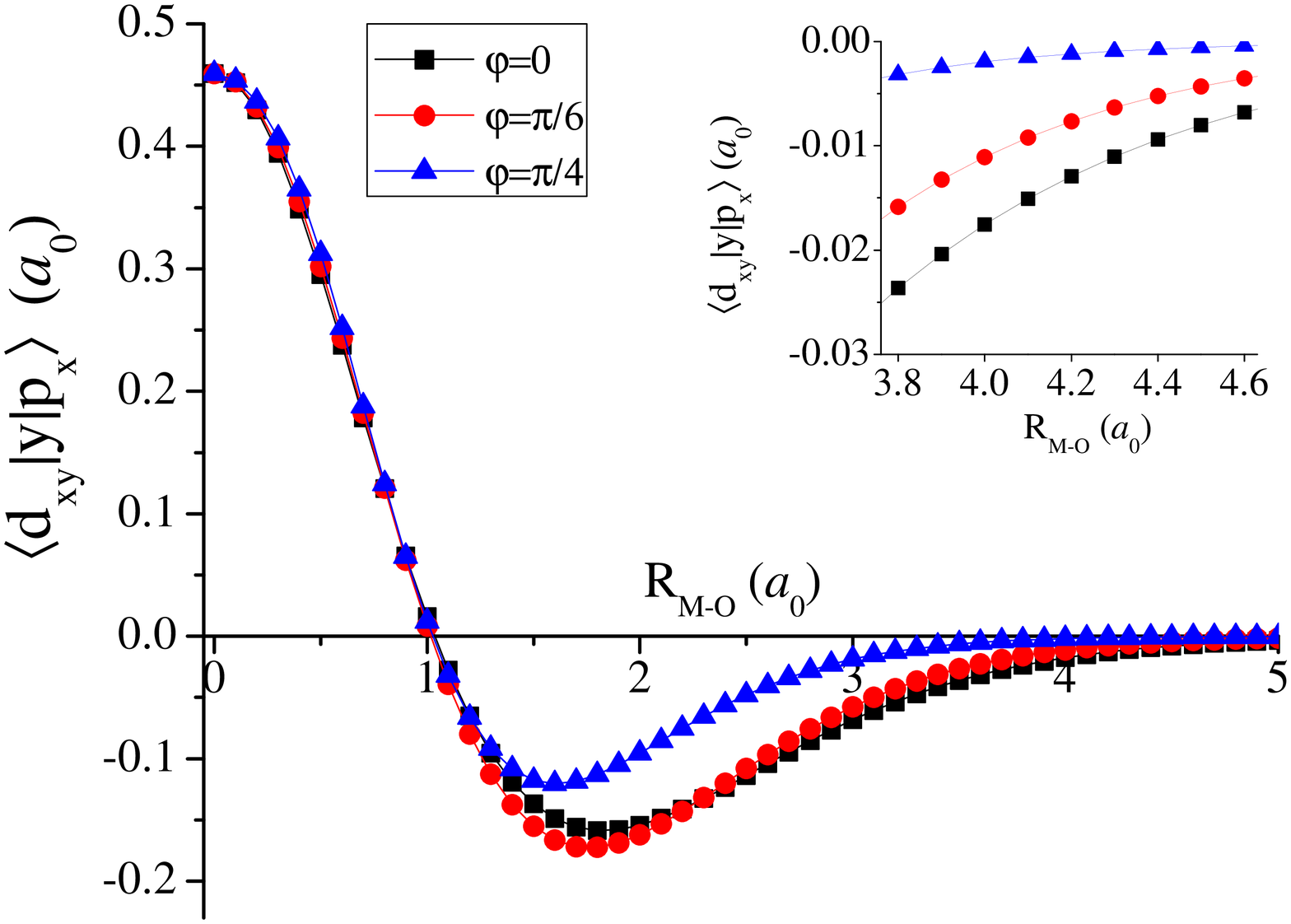}
\includegraphics[width=7cm]{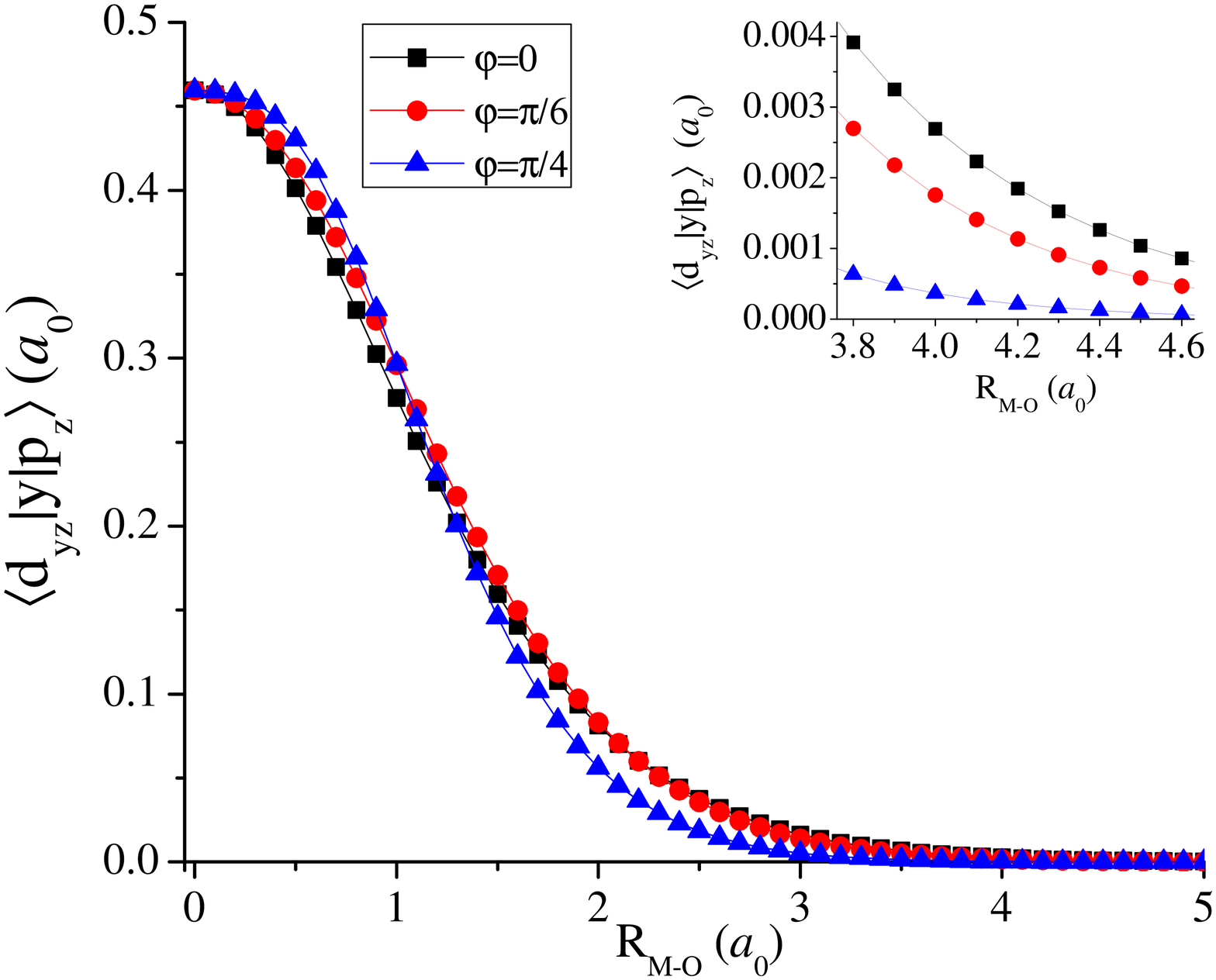}
\includegraphics[width=7cm]{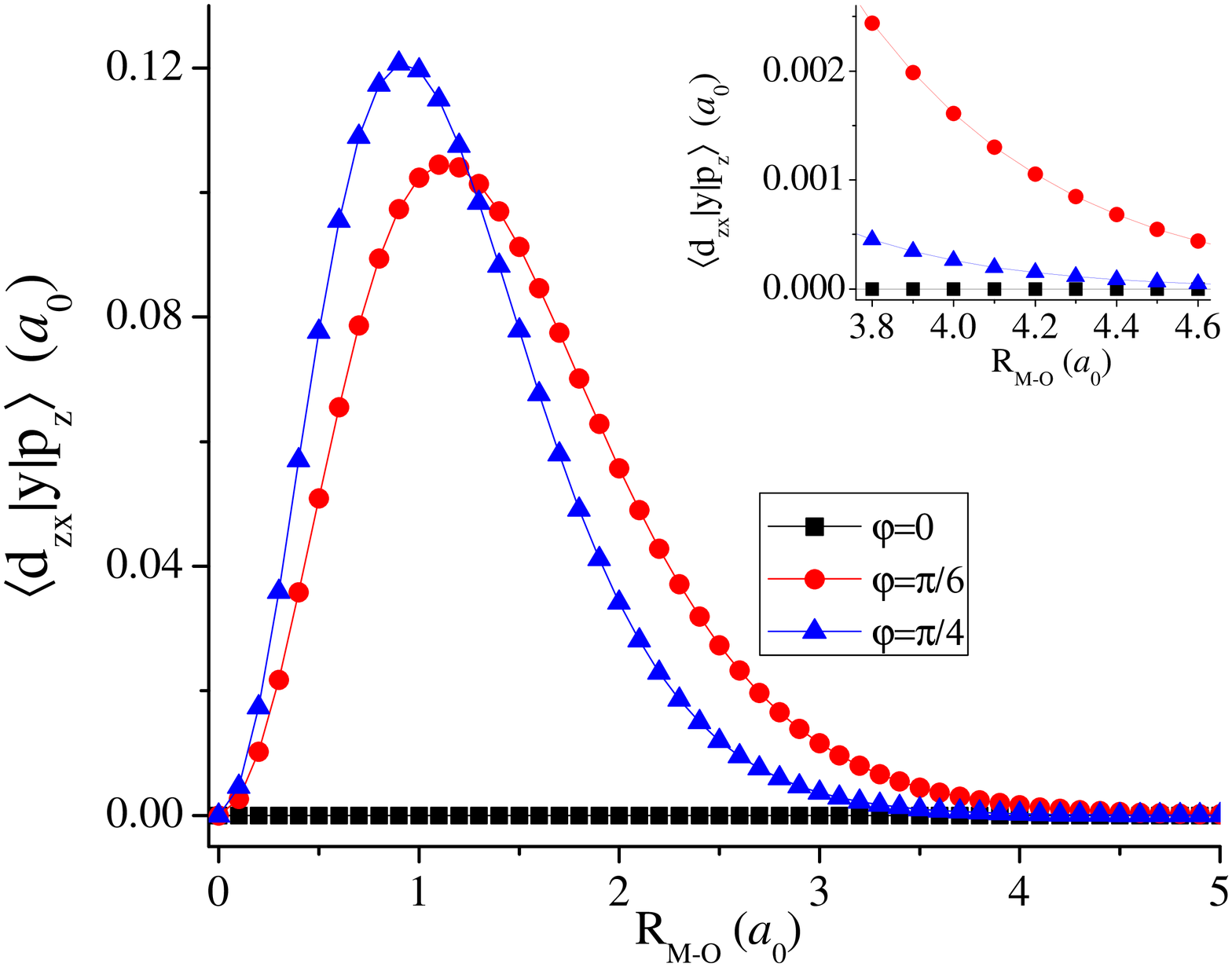}
\includegraphics[width=7cm]{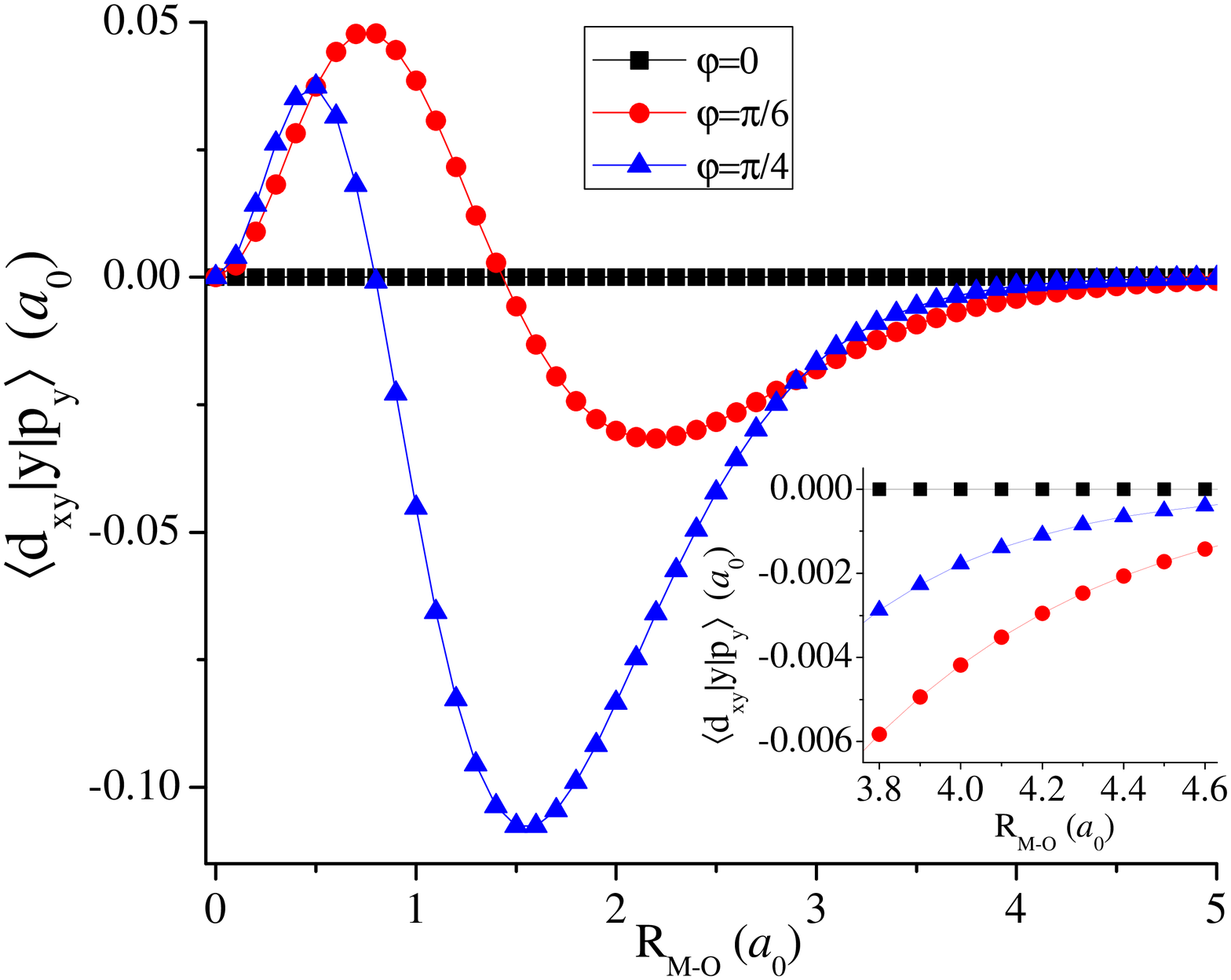}
\vspace{-0.3cm}
\end{center}
\caption{Two-site dipole integrals $I^{y}_{\mu,\nu}$ against M-O separation R$_{\text{M-O}}$ with various bond bending angle.
The inset in each figure shows the zoom in view of the integrals near typical M-O separation (i.e., 4$a_{0}$, $a_{0}$ is the Bohr radius).}
\label{fig:Iy}
\end{figure}

\begin{figure}[hptb]
\begin{center}
\centering
\includegraphics[width=7.0cm]{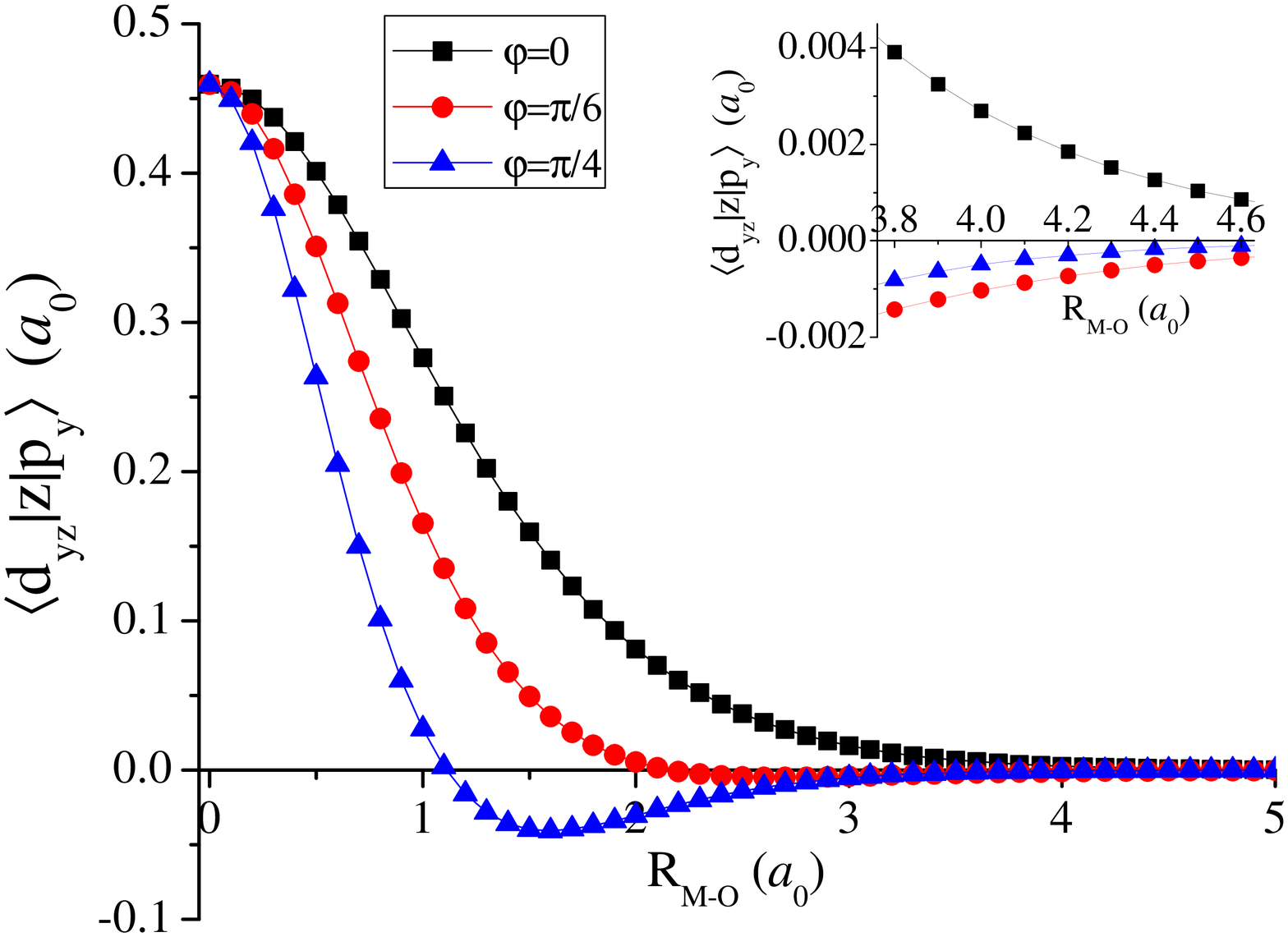}
\includegraphics[width=7.0cm]{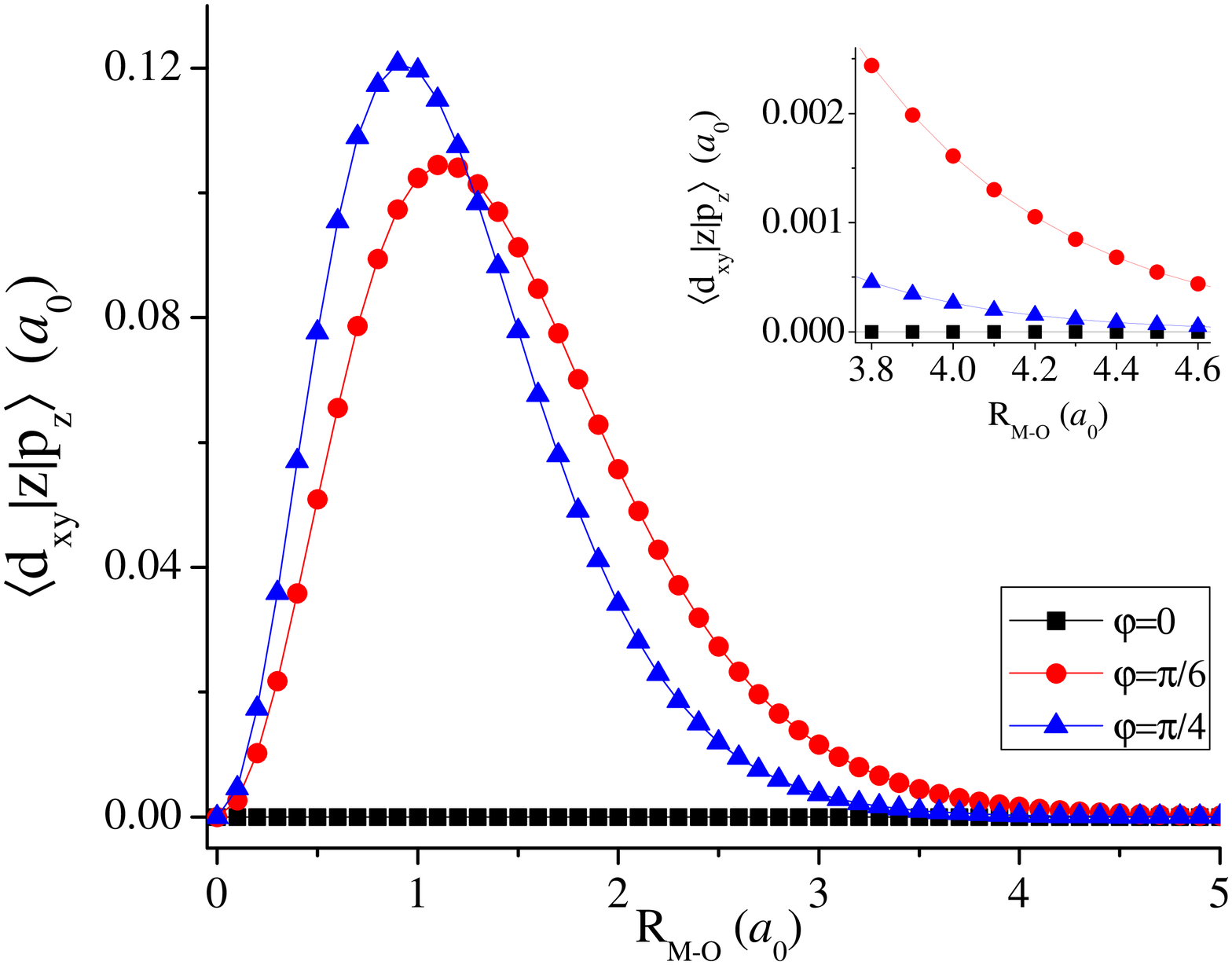}
\includegraphics[width=7.0cm]{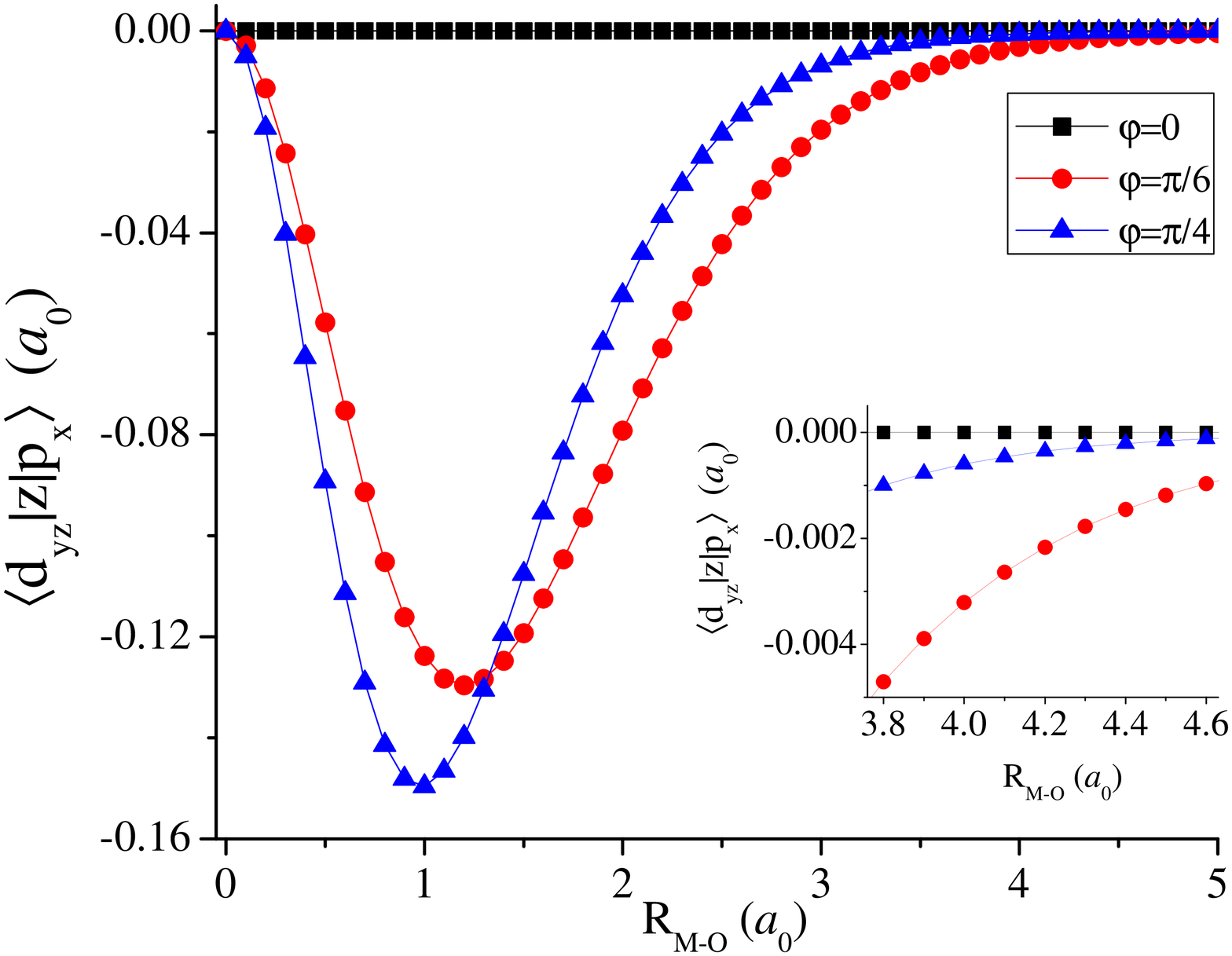}
\vspace{-0.3cm}
\end{center}
\caption{Two-site dipole integrals $I^{z}_{\mu,\nu}$ against M-O separation R$_{\text{M-O}}$ with various bond bending angle.
The inset in each figure shows the zoom in view of the integrals near typical M-O separation (i.e., 4$a_{0}$).}
\label{fig:Iz}
\end{figure}

It turns out that relevant $I^{\alpha}_{\mu,\nu}(\vec{a})$'s in the main text
include $I^{z}_{yz,y}$, $I^{z}_{xy,z}$, $I^{z}_{yz,x}$, $I^{y}_{xy,x}$, $I^{y}_{yz,z}$, $I^{y}_{zx,z}$, $I^{y}_{xy,y}$.
We demostrate the numerical evaluation for these integrals in Fig.\ref{fig:Iy} and Fig.\ref{fig:Iz} respectively.
R$_{\mathrm{M-O}}$ is the M-O bond distance along the M-M chain. In this calculation, we have taken the hydrogenlike radial-wave functions
and the Clementi-Raimondi effective charges $Z_{\mathrm{O}}^{\mathrm{eff}}=4.45$ and $Z_{\mathrm{M}}^{\mathrm{eff}}=10.53$
for O$^{2-}$ and Mn$^{3+}$ ions respectively\cite{CR-1}.
It is shown that, for typical M-O separation (i.e., $\text{R}_{\mathrm{M-O}} \sim 4a_{0}$, here $a_{0}$ is Bohr radius),
$I^{y}_{xy,x}$ is one order of magnitude larger than those of other integrals.
The separation between the M and O ions R$_{\mathrm{M-O}}$ not only yields
violent variance on the magnitude of the overlap dipole integrals but also may change their signs.

\end{document}